\documentclass{aa}  

\usepackage{graphicx}
\usepackage{txfonts}
\usepackage[switch]{lineno} 

\usepackage{natbib}
\bibpunct{(}{)}{;}{a}{}{,}

\usepackage{xcolor}
\usepackage[colorlinks=true, linkcolor=blue, citecolor=blue, urlcolor=blue]{hyperref}

\newcommand{\customcitecolor}[2]{%
    \begingroup
    \hypersetup{citecolor=#1}%
    \cite{#2}%
    \endgroup
}

\newcommand{\customcitecolornew}[2]{%
    \begingroup
    \hypersetup{citecolor=#1}%
    \citealt{#2}%
    \endgroup
}

\newcommand{\num}[1]{{\textcolor{blue}{#1}}}

\usepackage{orcidlink}
\begin{document} 

   \title{ZTF SN Ia DR2: An environmental study of Type Ia supernovae using host galaxy image decomposition}
   \author{ R. Senzel\inst{1,2}\orcidlink{0009-0002-0243-8199},
            K. Maguire\inst{1}\orcidlink{0000-0002-9770-3508},
            U. Burgaz \inst{1}\orcidlink{0000-0003-0126-3999},
            G. Dimitriadis\inst{1}\orcidlink{0000-0001-9494-179X},
            M.~Rigault\inst{3}\orcidlink{0000-0002-8121-2560},
            A.~Goobar\inst{4}\orcidlink{0000-0002-4163-4996},
            J.~Johansson\inst{4}\orcidlink{0000-0001-5975-290X},
            M.~Smith\inst{5}\orcidlink{0000-0002-3321-1432},
            M.~Deckers\inst{1}\orcidlink{0000-0001-8857-9843},
            L.~Galbany\inst{6,7}\orcidlink{0000-0002-1296-6887},
            M.~Ginolin\inst{3}\orcidlink{0009-0004-5311-9301},
            L.~Harvey\inst{1}\orcidlink{0000-0003-3393-9383},
            Y.-L.~Kim\inst{5}\orcidlink{0000-0002-1031-0796},
            T.~E.~Muller-Bravo\inst{6,7}\orcidlink{0000-0003-3939-7167},
            P.~Nugent\inst{8,9}\orcidlink{0000-0002-3389-0586},
            P.~Rosnet\inst{10}\orcidlink{0000-0002-6099-7565},
            J.~Sollerman\inst{11}\orcidlink{0000-0003-1546-6615},
            J.~H.~Terwel\inst{1,12}\orcidlink{0000-0001-9834-3439},
            R.~R.~Laher\inst{13}\orcidlink{0000-0003-2451-5482},
            D.~Reiley\inst{14},
            B.~Rusholme\inst{13}\orcidlink{0000-0001-7648-4142}
    }
    
   \institute{School of Physics, Trinity College Dublin, The University of Dublin, Dublin 2, Ireland\\\
              \email{senzelr@tcd.ie} }

    \institute{School of Physics, Trinity College Dublin, College            Green, Dublin 2, Ireland\\
            \email{senzelr@tcd.ie}
            \and Max-Planck-Institut f{\"u}r Radioastronomie, Auf dem H{\"u}gel 69, D-53121 Bonn, Germany
             \and Universite Claude Bernard Lyon 1, CNRS, IP2I Lyon / IN2P3, IMR 5822, F-69622 Villeurbanne, France\ 
            \and  The Oskar Klein Centre, Department of Physics, AlbaNova, Stockholm University, SE-106 91 Stockholm , Sweden\
            \and Department of Physics, Lancaster University, Lancs LA1 4YB, UK\
            \and Institute of Space Sciences (ICE-CSIC), Campus UAB, Carrer de Can Magrans, s/n, E-08193 Barcelona, Spain\ 
            \and Institut d'Estudis Espacials de Catalunya (IEEC), 08860 Castelldefels (Barcelona), Spain\ 
            \and Lawrence Berkeley National Laboratory, 1 Cyclotron Road MS 50B-4206, Berkeley, CA, 94720, USA \
            \and Department of Astronomy, University of California, Berkeley, 501 Campbell Hall, Berkeley, CA 94720, USA\
            \and Université Clermont Auvergne, CNRS/IN2P3, LPCA, F-63000 Clermont-Ferrand, France\
            \and The Oskar Klein Centre, Department of Astronomy,
            AlbaNova, Stockholm University, SE-106 91 Stockholm , Sweden \
            \and Nordic Optical Telescope, Rambla Jos\'e Ana Fern\'andez P\'erez 7, ES-38711 Bre\~na Baja, Spain\
            \and IPAC, California Institute of Technology, 1200 E. California Blvd, Pasadena, CA 91125, USA\
            \and Caltech Optical Observatories, California Institute of Technology, Pasadena, CA 91125, USA\
            }
              
 \titlerunning{ZTF SN Ia DR2 - Host Galaxy Decomposition}
 \authorrunning{R. Senzel et al.}
 \date{Received <date>; accepted <date>}

  \abstract
  {The second data release of Type Ia supernovae (SNe Ia) observed by the Zwicky Transient Facility has provided a homogeneous sample of 3628 SNe Ia with photometric and spectral information. This unprecedented  sample size enables us to better explore our currently tentative understanding of the dependence of host environment on SN Ia properties. In this paper, we make use of two-dimensional image decomposition to model the host galaxies of SNe Ia. We model elliptical galaxies as well as disk/spiral galaxies with or without central bulges and bars. This allows for the categorisation of SN Ia based on their morphological host environment, as well as the extraction of intrinsic galaxy properties corrected for both cosmological and atmospheric effects, through point-spread-function (PSF) convolution. We find that although this image decomposition technique leads to a significant bias towards elliptical galaxies in our final sample of processed galaxies, the overall results are still robust. By successfully modelling 728 host galaxies, we find that the photometric properties of SNe Ia found in disks and in elliptical galaxies, correlate fundamentally differently with their host environment. We identified strong linear relations between light-curve stretch and our model-derived galaxy colour for both the elliptical (16.8$\sigma$) and disk (5.1$\sigma$) subpopulations of SNe Ia. Lower stretch SNe Ia are found in redder environments, which we identify as an age/metallicity effect. Within the subpopulation of SNe Ia found in disk containing galaxies, we find a significant linear trend (6.1$\sigma$) between light-curve stretch and model-derived local $r$-band surface brightness, which we link to the age/metallicity gradients found in disk galaxies. SN Ia colour shows little correlation with host environment as seen in the literature. We do identify a possible dust effect in our model-derived surface brightness (3.3$\sigma$), for SNe Ia in disk galaxies. }
   \keywords{galaxies : fundamental parameters --
                supernovae: general --
                techniques: image processing
               }

   \maketitle
%

\section{Introduction}

Type Ia supernovae (SNe Ia) are fundamental tools in modern astrophysics and cosmology research. With their tremendous luminosities, SNe Ia act as standardisable candles \citep{phillips1993absolute, hamuy1995hubble, riess1996precise, tripp1998}, allowing us to probe vast distances in our Universe. Most famously, SNe Ia were used to discover the accelerated expansion of the Universe and the existence of a cosmological constant \citep{riess1998, perlmutter1999}. In the decades since, they have been used to improve the measurement of the dark energy equation-of-state parameter \textit{w}, as well as the local Hubble constant $H_0$, to ever-increasing precision \citep{garnavich1998supernova, perlmutter1999constraining, brout2019, riess2022}. However, major issues with our understanding of cosmology and the nature of the Universe are still present, such as the fundamental origin of dark energy \citep{abbott2024dark} as well as the so-called Hubble Tension, whereby the Hubble constant determined by SNe Ia in the local universe is incompatible with the measurement derived from the early universe at $>5\sigma$ \citep{planck, hubbletension, freedman2021_h0tension,riess2022}. If this disagreement is not due to (multiple) systematic uncertainties, this would be a sign of new unknown physics. It has become increasingly apparent that the physical nature of SNe Ia, how they are standardised, and any systematic uncertainties present must be better understood.

SNe Ia are widely believed to be the thermonuclear explosion of a carbon-oxygen white dwarf (C/O WD) in a binary system \citep{hoyle1960, maoz2014}. However, the exact nature of the companion star, the mechanism by which the white dwarf explodes, and how it relates to the observed diversity of SNe Ia, has not been solved \citep{liu2023}. A popular model is the single degenerate model \citep{sd}, consisting of a C/O WD with a non-degenerate companion star, such as a main-sequence or red giant star. The white dwarf accretes material from its companion star, increasing its mass. As the white dwarf's mass approaches the theoretical maximum \cite{chandrasekhar1935} mass ($\sim 1.4~M_{\odot}$), central densities overcome the electron degeneracy, triggering a runaway thermonuclear reaction.  The second main companion scenario (or progenitor channel), is the double degenerate model \citep{dd}, consisting of two white dwarfs. This system can trigger an SN Ia either by a direct merger (due to orbital decay via gravitational wave emission), or by accreting helium from one white dwarf onto the other, most likely resulting in a double-detonation \citep{nomoto1982, greggio2005, fink2010, woosley2011,Shen2018}.  In this scenario, the companion star donates sufficient amounts of helium, resulting in a helium detonation on the surface of the white dwarf, providing the necessary increase to the white dwarf's central density, triggering a sub-Chandrasekhar mass explosion.

An important avenue to further our understanding of SNe Ia is by studying their environments, i.e.~their host galaxies. In modern cosmology, there is a puzzling dependence of the corrected Hubble residuals (remaining scatter after standardisation) on the SN Ia host galaxy stellar mass, whereby higher mass galaxies contain SNe Ia that standardise to brighter luminosities \citep{sullivan2010, kelly2010, lampeitl2010effect, kim2019, smith2020, rigault2020, briday2022}. This is a significant source of systematic uncertainty and is commonly corrected for in modern studies via a mass-step \citep{betoule2014, scolnic2018, brout2022}.  Several studies have shown that galaxy colour also contains a step, and may be an overall better tracer of the correlation between SNe Ia and their host galaxy \citep{rigault2013, roman2018, rigault2020, kelsey2021}. Galactic properties, such as mass and colour, are known to correlate with stellar population age, metallicity, star-formation rate, dust content etc. \citep{massmetal}. However, it is unclear what the dominant physical driving factor behind the observed SN Ia mass-step is, necessitating a more physical understanding of the correlation between SN Ia properties and their host galaxies \citep{brout2021, briday2022}.

The impressive resolution and observing depths of modern (all-sky) imaging surveys have allowed us to study the structures of galaxies at vast distances. Galaxy morphologies are very complex, with no two galaxies being identical.  In classic galaxy formation, disks are young, dusty, and star-forming features, while bulges contain old populations of stars, with higher metallicity and very little dust and star-formation. Visual classification is usually performed by recognising common structural features within galaxies such as disks, bulges, bars, spiral arms etc. \cite[see][for a review on galaxy formation and structure]{benson2010galaxy, conselice2014evolution}. In order to extract structural information from these galaxy images, many `Bulge+Disk' (+other component) decomposition codes have been created, that fit two-dimensional (2D) surface brightness models to galaxy images, such as GALFIT \citep{galfit} and GIM2D \citep{gim2d}. These codes allow for easy extraction of useful structural information from galaxy images, while also benefiting from robust correctional methods for atmospheric and instrumental smearing effects. 

In this paper, we make use of the second data release of SNe Ia from the Zwicky Transient Facility (ZTF) \citep{bellm2019zwicky, graham2019zwicky} between 2018 and 2020 (`ZTF DR2',  \customcitecolornew{blue}{Rigault2024a}) and perform automated image decomposition of their host galaxies into bulge, disk, and bar components. This allows us to link the SN position to a specific stellar population and provide insights into the possible SN Ia progenitor channels (different companion stars are more common in certain environmental types, e.g.~high-mass main-sequence stars are found in young disks), as well as the potential effects of metallicity, age and dust extinction on SN properties.  This analysis uses a custom, gradient-descent based code\footnote{\url{https://github.com/RobertSenzel/ZTF_host_decomp}}, designed to perform fully automated 2D image decomposition for nearby (up to a redshift of $z\sim 0.15$) ground-based galaxy images. The code takes as input \textit{g}- and \textit{r}-band images from the Dark Energy Spectroscopic Instrument Legacy Imaging Surveys  \citep[DESI-LS,][]{desi}, and generates a 2D surface brightness model of the galaxy without any user input. The code handles external source masking (Section~\ref{sec:mask}), PSF convolution during fitting (Section~\ref{sec:psf}), initial condition estimation (Section~\ref{sec:init}), cosmological corrections (i.e. K-corrections, Section~\ref{sec:cosmo}) and morphological classification (Section~\ref{sec:edg_on}). Throughout this paper, the 2018 Planck Collaboration results \citep{planck} are used for cosmological calculations, made available by \texttt{AstroPy} \citep{astropy}.

\section{Data}
\label{sec:data}

In this section, we describe the SN Ia and host galaxy image data used in our analysis. In Section \ref{sec:desi}, we define the images used for analysing the host galaxies, while in Section \ref{sec:preprocess}, we detail the image preprocessing. In Section \ref{sec:salt}, we introduce the ZTF SN Ia sample,  quality cuts that are applied to the sample, and detail the parameters measured from the light curves.

\subsection{DESI Legacy Imaging Surveys}
\label{sec:desi}
\begin{figure*}
    \centering
	\includegraphics[width=17cm]{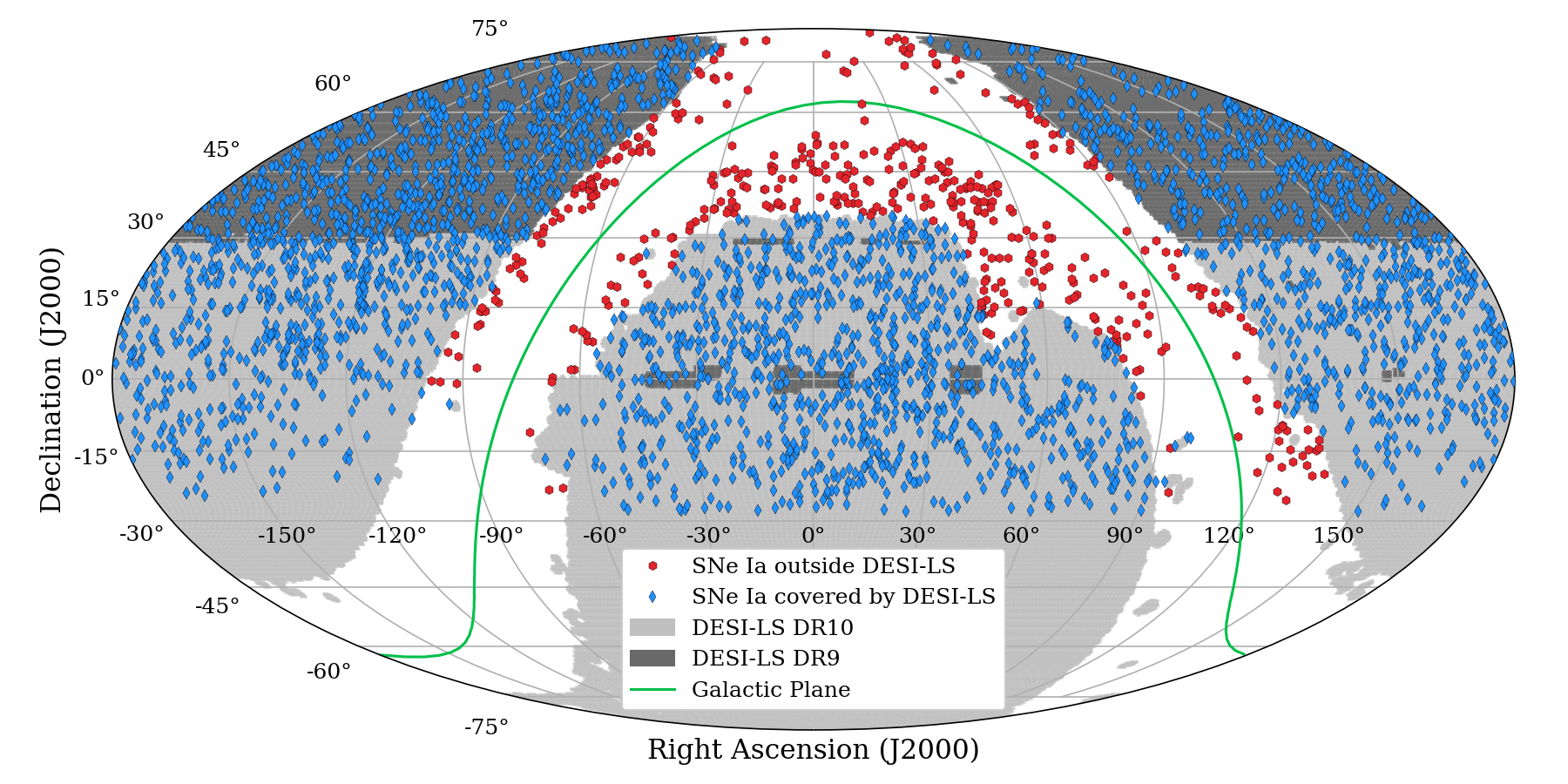}
    \caption {DESI-LS sky coverage shown using a Mollweide projection, with DR10 in light-grey and DR9 in dark-grey blue. The blue points represent covered ZTF Cosmo DR2 SNe Ia (83\%), while the red points are SNe Ia outside the sky coverage. The green line marks the Galactic plane.} 
    \label{fig:skymap}
\end{figure*}

The galaxy images of the SN Ia sample used in this study come from DESI-LS \citep{desi}. We make use of $g$ and $r$-band images, due to these bands having the deepest available observing depths (AB magnitude system). The DESI-LS consists of three separate surveys, with the joint goal of supplying optical and infrared imagery of 14,000 deg$^2$ of extra-galactic sky for the DESI project.  The Beijing-Arizona Sky Survey (BASS) uses the 2.3m Bok telescope to observe in $g$ and $r$-bands, while the Dark Energy Camera (DECam) Legacy Survey uses the 4m Blanco telescope in Chile, to capture $grz$ imagery. The third survey, the Mayall $z$-band Legacy Survey, only observes in the infrared and is not used in this paper. The latest two data releases DR9 and DR10 are used in this paper, which when combined, provide galaxy images for 83\% of the SNe Ia in the ZTF DR2, as seen in Fig.~\ref{fig:skymap}. The missing SNe Ia are due to the DESI sky coverage having a larger cut-out for the Milky Way, relative to the ZTF sky coverage. While the observing period for the ZTF DR2 overlaps with DESI-LS DR10, contamination of the galaxy images by SN light is not a concern, as this study uses fitted galaxy models to extract information, rather than using the actual pixels at the SN location.

\subsection{Preprocessing of DESI-LS images}
\label{sec:preprocess}
Each galaxy image is obtained by querying DESI-LS image cutouts from their sky-viewer\footnote{\url{http://legacysurvey.org/viewer}}. The coadded images (image stacks) are used, which are fixed at a pixel scale of 0.262 arcsec/pixel. The sky level has already been subtracted from these images. The centre of the image is taken from the matched host galaxy information for each ZTF DR2 SN Ia \num{Smith et al. (in prep.)}, while the size of the image is set to be a $70 \times 70$ $\mathrm{kpc^2}$ aperture, calculated using the ZTF SN Ia redshifts. The maximum image size is limited to 600 pixels, corresponding to a 70 kpc aperture at $z\sim0.02$. Consequently, galaxies closer than this redshift have apertures smaller than 70 kpc, and hence may have a larger angular diameter than the cutout image. The quality cuts detailed in Section~\ref{sec:iso_cut} ensure that only galaxies smaller than their cutout image are fitted. 

The DESI-LS cutout service returns an image in flux space (in units of nanomaggie) as well as a corresponding inverse-variance image. The header information contains the sky brick ID, which is used to obtain the full-width-half-maximum (FWHM, in units of arcsec) of the PSF and the $5\sigma$ detection depths for the $g$ and $r$-bands. Images that lie on brick edges are returned as multiple images and were stitched together (totalling approximately 4\% of the sample). The PSF FWHM and the $5\sigma$ detection depths are estimated by averaging the values from each image. Images that lie on three or more bricks (and hence returned as three or more separate images) were discarded, as there were only nine such events (0.3\%). 

The flux image, F (in units of nanomaggie), is converted to apparent surface brightness, $\mu$ (in units of mag/arcsec$^2$), using the equation,
\begin{equation}
    \mu = \mathrm{ZP} - 2.5\mathrm{log}_{10}\left(\frac{F}{\mathrm{scale}^2}\right) - \mathrm{A}_{\mathrm{MW}}
	\label{eq:sb_eq}
\end{equation}
where the zero-point $(\mathrm{ZP})=22.5$ mag and the scale $=0.262$ arcsec/pixel, are both defined by DESI-LS. A$_\mathrm{MW}$ is the Milky Way dust extinction, calculated using a total-to-selective extinction ratio, R$_\mathrm{V}$ of 3.1 and the extinction model from \cite{dustmw}. Surface brightness is commonly used in galaxy studies. It is a distance independent quantity, related to the density of stars at any particular location within a galaxy. The $1/d^2$ dependence of flux is cancelled out by dividing in the pixel scale. With increasing distance, the resolution of the galaxy image (i.e. how many pixels per galaxy) decreases, but the measured surface brightness remains constant.

A shortcoming of Equation~\ref{eq:sb_eq} is that it cannot represent non-detections, i.e. flux values less than zero. The image processing described in subsequent sections requires 2D images containing real numbers. A hyperbolic transformation is performed as described in \cite{lupton99}, that generates reasonable magnitudes for negative flux values, while leaving high signal-to-noise ratio magnitudes virtually unchanged. This transformation replaces the logarithm with:
\begin{equation}
    -2.5\mathrm{log}_{10}\left(F\right) = -\frac{2.5}{\ln(10)} \left[\mathrm{asinh}\left(\frac{F}{2b_\mathrm{s}}\right) + \ln(b_\mathrm{s})\right]
	\label{eq:arcsin}
\end{equation}
where $b_\mathrm{s}$ is a softening parameter, taken from \cite{sloan}, as $0.9\times10^{-10}$ in the $g$ band and $1.2\times10^{-10}$ in the $r$ band. Using this transformation for high signal-to-noise values, the dispersion in magnitude from equation~\ref{eq:sb_eq} is on the order of $10^{-6}$ mag.

\subsection{ZTF Cosmo DR2}
\label{sec:salt}
ZTF \citep{masci2018zwicky,bellm2019zwicky,graham2019zwicky,dekany2020zwicky} is a state-of-the-art optical survey that scans the entire Northern sky every two-three nights ($g$, $r$ and $i$ band).  Since it began operation in 2018, it has detected a very large sample of SNe Ia. The ZTF SN Ia cosmology project \cite[see][for the first data release]{ztfdr1}, aims to use these SNe Ia as distance indicators to further our understanding of cosmology. This paper makes use of the second data release (ZTF DR2), containing 3628 spectroscopically classified SNe Ia, from the first three years of operation (2018-2020). For a full overview of the dataset, including technical details on photometric and spectral data, redshift measurements, spectral classifications, and host galaxy matching, see \customcitecolor{blue}{Rigault2024a} and \num{Smith et al. (in prep.)}.

The redshifts used for this sample come from three sources. Public galaxy catalogues (such as NASA/IPAC Extragalactic Database) are used for 60.6\% of the sample, of which 71\% comes from DESI through the Mosthost program \citep{soumagnac2024most}. These redshifts, along with a further 8.8\% coming from visible galaxy emission lines in the ZTF SN Ia spectra, are known as `galaxy redshifts' and have a typical precision of $\Delta z \le 10^{-4}$. The remaining 30\% of the sample derive their redshift measurements from the SNID SN template matching programme \citep{blondin2007determining}. A comparison between galaxy- and SNID-redshifts was carried out by \num{Smith et al. (in prep.)}, who find a mean SNID-redshift precision of $\Delta z \approx 10^{-3}$.

This paper mainly uses originally derived galaxy parameters to investigate correlations with SNe Ia. However, we also make use of the computed host galaxy stellar masses, provided in the ZTF SN Ia DR2. As described in  \customcitecolor{blue}{Rigault2024a}, the host stellar masses are computed by spectral-energy-distribution fitting of global photometry using the P{\`E}GASE2 galaxy spectral templates  \citep{le2002photometric}. The global photometry is derived from the \texttt{HostPhot} package \citep{muller2022}, using public Pan-STARRS1 DR2 photometry.

The ZTF SN Ia DR2 contains detailed spectroscopic classification of the SN Ia sample. These can be grouped into five subclasses, `normal SNe Ia', `91bg-like’ \citep{filippenko1992_91bg}, ‘91T-like’  \citep{filippenko1992_91t}, `peculiar SNe Ia' and `unclear SNe Ia'. The `91T-like' subclass also contains ‘99aa-like’ events, as they are generally considered transitional SNe Ia between ‘91T-like’ and ‘normal SN Ia’, with very similar photometric properties \citep{phillips2022, 91t}.  The `peculiar' subclass contains all identified subclasses that do not fall into these initial ones, e.g.~SNe, Iax, `03fg-like', see  \customcitecolor{blue}{Burgaz2024}. The last subclass of `unclear' are the SNe Ia that are classified as SNe Ia, but their subclassification cannot be determined from their spectra. In this analysis, we remove these last two subclasses (`peculiar' and `unclear'), as they are not relevant for this analysis. The `peculiar' subclass is removed, as it is composed of rare subtypes with too few numbers to do any meaningful analysis of their host galaxy properties. The `unclear' subclass is removed, as it does not represent a physical subgroup of SNe Ia, but rather a sample of SNe Ia characterised by insufficient spectral information. This group can contain any of the SN Ia subclasses, with late-time spectra, or spectra with host contamination, where subclassification becomes difficult. These subgroups are removed after the galaxy fitting procedure and galaxy quality cuts. 

\begin{figure}[h]
	\includegraphics[width=\columnwidth]{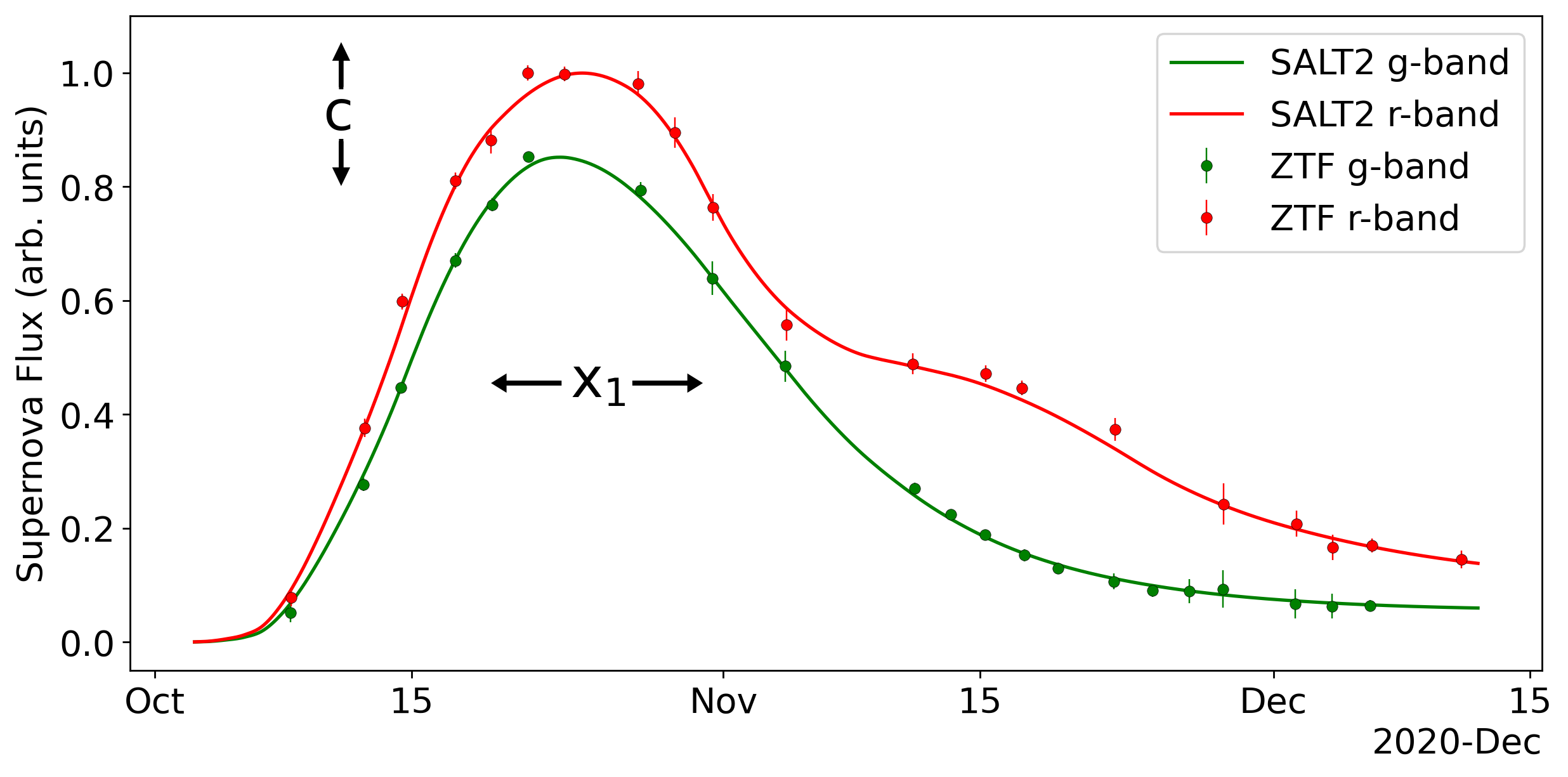}
    \caption {An example SN Ia rest-frame light curve from ZTF DR2, with a fitted SALT2 model. The $x_1$ parameter quantifies the stretch of the light curves, while  $c$ quantifies the colour. The black arrows are for illustrative purposes only, they do not describe how these parameters are calculated. } 
    \label{fig:salt_img}
\end{figure}

The photometric parameters used in this study are those produced by the SALT2 light curve model \citep{guy2005salt, salt2, betoule2014, taylor2021}, namely a parameter related to the width of the light curve $x_1$ and one related to the colour $c$, see Fig.~\ref{fig:salt_img}. These are the commonly used parameters for the standardisation of SNe Ia used in modern precision cosmology. The ZTF SN Ia photometry and light-curve quality cuts are applied after the galaxy fitting procedure and galaxy quality cuts, as some figures do not require light-curve information. SNe Ia with satisfactory light-curve coverage are defined as having at least seven phase detections across two photometric bands, with at least two phase detections before and two after peak luminosity (same night detections in the same band, count as only one phase). We require the uncertainty in the SALT2 light curve width of $\Delta$x$_1 < 1$, the uncertainty in the colour of $\Delta c < 0.1$ and `fitprob' $>10^{-7}$ (this is a SALT2 specific parameter quantifying the probability of the light-curve being that of a SN Ia). For further discussion on the applied cuts, see \customcitecolor{blue}{Rigault2024a}. 

This paper, investigates the correlations between SN Ia light-curve properties and their host galaxies, with our main focus on non-peculiar SNe Ia that can be standardised for use in cosmology. For the ‘normal SN Ia’ and ‘91T-like’ SNe Ia, we apply the  SALT2 parameter cuts of $-3<x_1<2$ and $-0.2<c<0.5$. These are similar to the `Basic cuts' from \customcitecolor{blue}{Rigault2024a},. We raised the upper bound of the colour parameter from the more standard cut of $c<0.3$, to allow for an investigation into any potential effects of dust extinction. For the ‘91bg-like’ SNe Ia, we do not apply the `fitprob' cut as the SALT2 model is not optimised for ‘91bg-like’ events, and we restrict the SALT2 parameter ranges to be $-5<x_1<-1$ and $-0.2<c<0.8$.

This study does not concern itself with SN Ia rates or a volume limited sample, and hence we are lenient with the redshift cuts to maximise our sample size. No redshift cut is applied to the ‘normal SN Ia’, however due to the limitations of galaxy fitting at high redshift, the final sample only contains SNe Ia with $z < 0.15$. For our analysis in Section \ref{sec:Results_gen} concerning the SN Ia subgroups, we study the difference in host galaxy property distributions of the ‘91bg-like’ and ‘91T-like’ events, relative to each other and to the overall distribution of SNe Ia. As discussed in \num{Dimitriadis et al. (submitted)}, and \customcitecolor{blue}{Burgaz2024}, careful checking of the spectral subclasses was only performed up to a redshift of $z=0.06$.  It is important that we have the correct subclassifications for these rarer subtypes, hence we apply a redshift cut of $z < 0.06$ for these two subgroups. These redshift and light curve cuts are only applied after the galaxy fitting procedure, as the final sample of processed galaxies is studied first without SN dependent cuts.

\begin{figure*}
\sidecaption
\includegraphics[width=12cm]{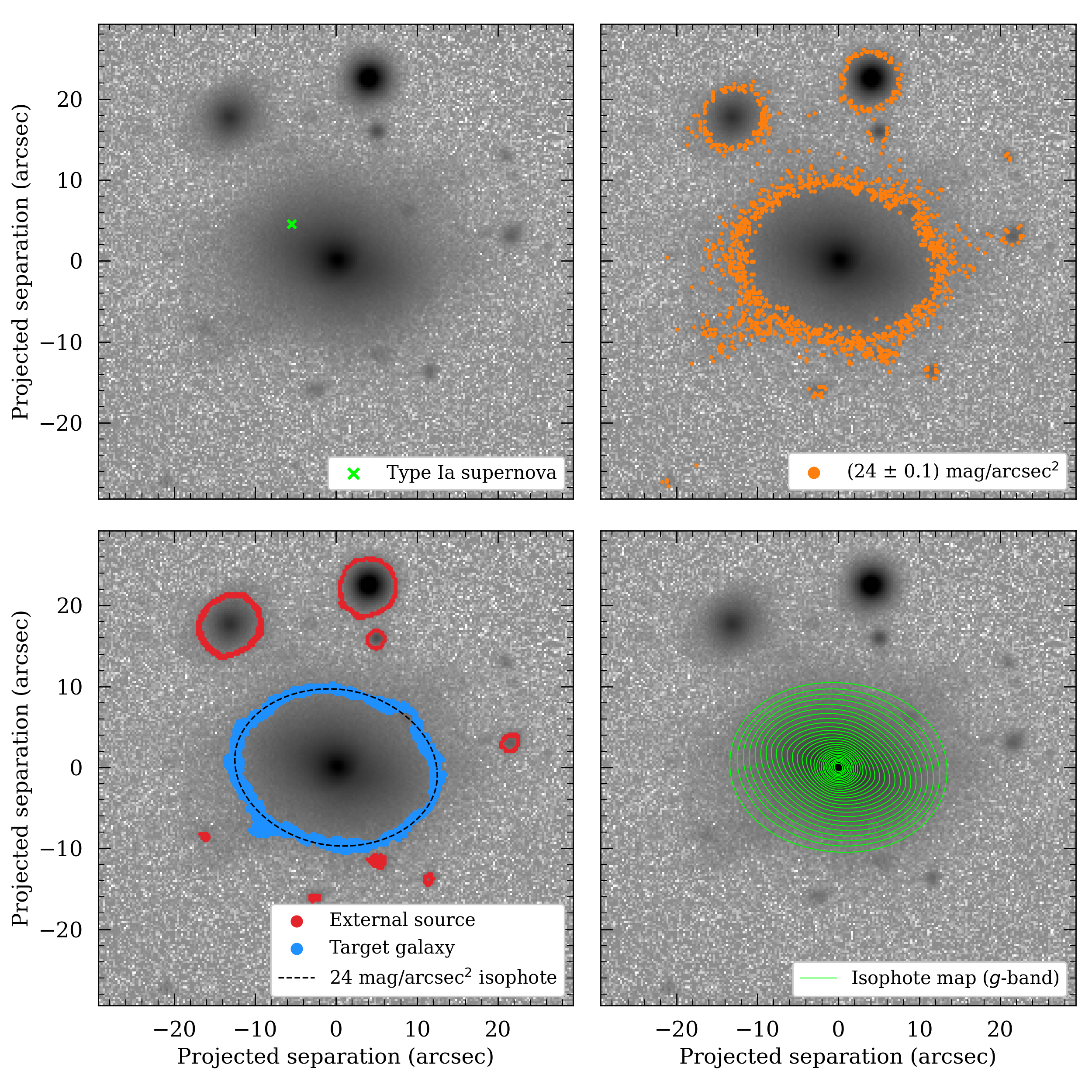}
\caption{\textbf{Top-left}: Example $g$-band DESI-LS image of an elliptical host galaxy, with neighbouring field sources. \textbf{Top-right}: The galaxy image sampled before mean-kernel convolution, at $24 \pm 0.1$ mag/arcsec$^2$. \textbf{Bottom-left}: The image sampled after mean-kernel convolution, with the target galaxy pixels highlighted in blue, external sources in red and a fitted 24 mag/arcsec$^2$ isophote in black. The target galaxy pixels are isolated by running a connected-component labelling algorithm. \textbf{Bottom-right} The constructed  $g$-band isophote map for this host galaxy, showing isophotes at every 0.2 mag/arcsec$^2$.}
\label{fig:contour}
\end{figure*}

\section{Constructing an isophote map}
\label{sec:construct_isophote}
This section details how isophote maps (contours of constant surface brightness) are created from DESI-LS images of SN Ia host galaxies. These maps are used in the galaxy fitting procedure, detailed in Section \ref{sec:galaxy_modelintro}. In Section \ref{sec:mask}, we describe how galaxy pixels are isolated from neighbouring sources. In Section \ref{sec:iso_fitting}, we describe the isophote map generation procedure, while the applied quality cuts are described in Section \ref{sec:iso_cut}.

\subsection{Isolation of galaxy pixels}
\label{sec:mask}
To isolate galaxy pixels from neighbouring sources, we developed a custom contour fitting algorithm (written in Python 3). After preprocessing of the DESI-LS image is complete (Section \ref{sec:preprocess}), a mean smoothing is applied to reduce noise. This consists of performing a numerical convolution to the image using a normalised $5\times5$ mean-kernel, in pixel space. The atmospheric PSF kernel size is nearly an order of magnitude larger, hence any information lost due to this smoothing is negligible (see Section~\ref{sec:psf}).  A two pixel strip is masked from all four edges to remove edge effects. To fit a contour for any given surface brightness value, the smoothed image is sampled for that value, within some range, e.g. $\mu = 24 \pm 0.1$ mag/arcsec$^2$. Fig.~\ref{fig:contour} shows the result of sampling the galaxy image before smoothing (top-right panel) and after smoothing (bottom-left panel).

This results in the individual sources within the image, forming isolated and connected regions. A connected-component labelling algorithm is applied from the Python image processing library, \texttt{scikit-image} \citep{scikit-image}. This algorithm labels each isolated region within the sampled pixels, allowing for easy differentiation between galaxy pixels with neighbouring sources, as seen in the bottom-left panel of Fig.~\ref{fig:contour}. The host galaxy pixels are determined by the labelled connected region whose geometric centre is closest to the image centre (defined by \num{Smith et al. in prep.}). 

\subsection{Isophote fitting}

\label{sec:iso_fitting}
Having isolated the desired galaxy pixels, we fit a superellipse to the pixels, generating an isophote. A superellipse (or Lamé curve) is a generalised form of an ellipse described by:
\begin{equation}
    \left|\frac{x}{a}\right|^{c_e} + \left|\frac{y}{b}\right|^{c_e} = 1
	\label{eq:se}
\end{equation}
where $a$ is the semi-major-axis and $b$ is the semi-minor-axis. The deviation from an ellipse is controlled by the superellipse index, $c_e$ ($c_e=2$ for an ellipse). Increasing $c_e$ above two increases the boxiness of the shape, while decreasing the value below two, flattens the edges towards a rhombus at $c_e=1$. This approach is common in `bulge+bar+disk' decomposition codes \citep{sell}, allowing for greater flexibility when fitting galaxy contours. The galaxy pixels are fitted for the semi-major-axis ($a$), semi-minor-axis ($b$), position angle ($\phi$), superellipse index ($c_e$) and the central coordinates ($x_c, y_c$), totalling six free parameters (the position angle is introduced by fitting in polar coordinates).

This process is performed on a range of surface brightness values in both the $g$ and $r$-bands. The outermost galaxy isophote is defined at the $5 \sigma$ depth for that particular image, with isophotes every 0.2 mag/arcsec$^2$, moving towards the centre. This process terminates when an isophote fit with less than 30 pixels is attempted, i.e. at least five pixels are required for each of the six free parameters in a superellipse. The width of the pixel sampling (e.g. $\pm$ 0.1 mag/arcsec$^2$  as shown in Fig.~\ref{fig:contour}), is optimised for every isophote fit, by minimizing the sum of the fractional uncertainty of the fitted superellipse parameters. This approach decreases the sampling width as far as possible without risking region fragmentation. This occurs when the sample width becomes too narrow and the galaxy pixels fragment into several different regions, preventing the connected-component labelling algorithm from extracting all the pixels associated with the galaxy. 

During the superellipse fitting, each pixel is weighted by its associated inverse-variance $\sigma^{-2}$ (nanomaggie$^{-2}$). These weights are converted to magnitude space using:
\begin{equation}
    \sigma_{mag} = \frac{2.5}{\ln(10)}\left|\frac{\sigma}{\mathrm{Flux}}\right|
	\label{eq:sigma}
\end{equation}
The uncertainty of pixels is naturally lower for brighter surface brightness (easier to detect), and hence biases the fit towards pixels with brighter values, e.g. fitting a superellipse to the range $24 \pm 0.1$ mag/arcsec$^2$, will result in an isophote closer to 23.9 mag/arcsec$^2$. To prevent this bias, the relation between pixel inverse-variance and surface brightness is divided out, resulting in a relative pixel uncertainty that can be used for isophote fitting.

This fitting procedure is used to generate $g$- and $r$-band isophote maps for all SN Ia host galaxies in the ZTF DR2 with DESI-LS images (2910 galaxies). An example isophote map is shown in the bottom-right panel of Fig.~\ref{fig:contour}. 

If the algorithm reaches a surface brightness value brighter than the galaxy centre, the process terminates, as there are no more pixels to fit. However, if an external source exists within the image, that is brighter than the target galaxy (e.g. Milky Way star), then the algorithm will continue to detect pixels and start producing isophotes from the external source, rather than terminating. To ensure that only isophotes belonging to the target galaxy are included in the isophote map, we require the centre of each isophote to be within ten pixels of the image centre. This cut removes on average 12\% of the generated isophotes for each isophote map.  With this cut applied, the central coordinate of the galaxy is calculated as the median of all isophote centres. This remains fixed during the galaxy fitting procedure, detailed in Section \ref{sec:galaxy_modelintro}.

\subsection{Quality cuts on isophote map creation}
\label{sec:iso_cut}

 \begin{table}
	\centering
	\caption{A breakdown of the isophote map fitting results for SN Ia host galaxies in ZTF Cosmo DR2.}
	\label{tab:iso_map_cuts}
	\begin{tabular}{l c c c}
        \hline\\[-0.5em]
		Cut & Count& Removed & \% Removed\\[0.15em]
		\hline\\[-0.5em]
        Full sample & 3628 & -& -\\[0.15em]
        \hline\\[-0.5em]
        In DESI-LS footprint & 3026 & 602 & 16.6\\[0.15em]
        Has identified galaxy & 2910 & 116 & 3.8\\[0.15em]
        Isophote map cut 1 & 2072 & 838 & 28.8\\[0.15em]
        Isophote map cut 2 & 2063 & 9 & 0.4\\[0.15em]
        Isophote map cut 3 & 1995 & 68 & 3.3\\[0.15em]
		\hline\\[-0.5em]
	\end{tabular}
\end{table}

Once an isophote map is generated, a number of quality cuts are applied to ensure sufficient coverage of the galaxy, before a galaxy decomposition is attempted. The breakdown of the quality cuts applied is shown in Table \ref{tab:iso_map_cuts}. As seen in Fig.~\ref{fig:skymap}, 17\% (602 SNe Ia) of our sample fall outside the DESI-LS coverage.  From the host galaxies with DESI-LS image, 4\% (116 SNe Ia) have no identified  host galaxy. These are SNe Ia with no host galaxy within 100 kpc (see \num{Smith et al. in prep.}) or due to these host galaxies covering less than 30 pixels on the image (galaxies with high redshift and/or low mass).

 The first isophote map cut requires the isophote map to have at least ten isophotes in both $g$ and $r$-bands, to ensure that the isophote map has enough coverage of the target galaxy, such that enough information (degrees of freedom) is available to fit all four of the galaxy models used in this paper (Table~\ref{tab:gal_model}). Each isophote provides four degrees of freedom (centre not included), therefore,  ten isophotes in both the $g$- and $r$-bands provide 80 degrees of freedom. The largest model (`bulge+bar+disk') has 17 free parameters, hence setting the cut at ten isophotes provides at least $80/17\sim5$ degrees of freedom per free parameter. 
 
 The second isophote map cut requires the difference in number of isophotes between $g$ and $r$-band to be less than ten. This ensures that equal coverage is available in both $g$ and $r$-bands, otherwise the fitted galaxy colour (see Section~\ref{sec:scc}) would be biased by whichever band had more coverage. For reference, the number of isophotes in either $g$ or $r$-band, ranges from ten up to an observed maximum of 40 isophotes. For a galaxy to have a difference in isophote number higher than ten (same order of magnitude as number of isophotes), it likely means there is an issue with one of the images (e.g. image artefacts, large difference in atmospheric PSF/ sky level, etc.). 
 
 The third isophote map cut requires the dimmest isophote in both the $g$- and $r$-bands to be dimmer than 23 mag/arcsec$^2$. This is applied to ensure that there are isophotes covering the outer regions of a galaxy, preventing the galaxy fitting routine from fitting only the bulge and not the disk. This cut will also remove galaxies which do not fit within the image aperture size. The value 23 mag/arcsec$^2$ is chosen as this surface brightness value is beyond the 5$\sigma$ observational depth ($g$ and $r$-bands) in both DESI-LS DR9 and DR10.
 
A discussion on how these cuts affect low mass and high redshift galaxies is presented in Section~\ref{sec:Results_gen}.  Following these quality cuts, 1995 successful isophote maps have been created with satisfactory coverage of the galaxy.

\section{Fitting a galaxy model}
\label{sec:galaxy_modelintro}

This section details how we model the 2D surface brightness of galaxies, starting from the isophote maps generated in Section \ref{sec:construct_isophote}. In Section~\ref{sec:models} we define the equations used to model the surface brightness of the various morphological components within galaxies. The fitting procedure is detailed in
Section~\ref{sec:gal_fit}, which is visually summarised in Fig.~\ref{fig:procedure}. Section~\ref{sec:psf} describes how the effects of atmospheric distortion, i.e.~point spread function (PSF) corrections, are removed from the galaxy image. In Section~\ref{sec:init}, the initial conditions are described, which are designed to maximise the probability of converging on a correct solution. Section~\ref{sec:cosmo} details the required cosmological corrections. Section~\ref{sec:edg_on} describes how the galaxy morphology is identified. The results of the galaxy fitting and morphology identification is broken-down in Section~\ref{sec:gal_fit_res}. Section~\ref{sec:sn_component} describes how a SN is identified with a galaxy component, while Section~\ref{sec:scc} defines how galaxy colour is calculated.

\subsection{Galaxy models}
\label{sec:models}
The galaxy profiles used in this study are the commonly used Sérsic profiles. These are 1D functions that relate surface brightness to the radial distance from the galaxy centre, \textit{R}. For a summary, see \cite{sersic}. The profile used for elliptical galaxies, bulges and bars is the general Sérsic profile:
\begin{equation}
    \mu(R) = \mu_e + \frac{2.5 b_n}{\ln(10)}\left[(R/R_e)^{1/n} -1 \right]
	\label{eq:sersic}
\end{equation}
where $R_e$ is the effective radius (radius containing half of the galaxy light), $\mu_e$ is the effective surface brightness (the surface brightness at $R_e$) and n is the Sérsic index, which controls the shape of the profile (higher n means more light concentrated near the centre). The geometric term $b_n$ is computed by numerically solving, $\Gamma(2n) = 2\gamma(2n,b_n)$, where $\Gamma$ and $\gamma$ are the complete and incomplete gamma functions respectively. Using Sterling's approximation, $b_n$ can be expressed as an asymptotic expansion \citep{sterling}, avoiding the need for expensive numerical solving,
\begin{equation}
    b_n = 2n - \frac{1}{3} + \frac{4}{405n} + \frac{46}{25515 n^2} + \mathrm{O}(n^{-3})
	\label{eq:sbn}
\end{equation}
 An exponential profile is used for galaxy disks, i.e. setting the Sérsic index to one,
\begin{equation}
    \mu(R) = \mu_0 + \frac{2.5}{\ln(10)}\left(\frac{R}{h}\right).
	\label{eq:disk}
\end{equation}
where $\mu_0$ is the central surface brightness and \textit{h} is the scale length. These can be transformed to effective surface brightness/radius using,
\begin{equation}
    \mu_e = \mu_0 + \frac{2.5b_n}{\ln(10)},
	\label{eq:conv_mu}
\end{equation}

\begin{equation}
    R_e = {b_n}^n h.
	\label{eq:conv_Re}
\end{equation}

This paper uses four main galaxy models, summarised in Table~\ref{tab:gal_model}. A single Sérsic profile is used for elliptical galaxies (early-type galaxies), labelled as the `Elliptical' model. For late-type galaxies, a single disk profile is used for a `Disk-only' galaxy model, a Sérsic and a disk profile are used for the `Bulge+Disk' model, and a disk with two Sérsic profiles are used for the `Bulge+Bar+Disk' model.

\begin{table} 
	\centering
	\caption{Overview of galaxy models used in this paper}
	\label{tab:gal_model}
	\begin{tabular}{l c c} 
        \hline\\[-0.5em]
		Model & Equation(s) & No. of Free parameters\\
		\hline\\[-0.5em]
		Elliptical &(\ref{eq:sersic}) & 7\\[0.15em]
        Disk-only & (\ref{eq:disk}) &6\\[0.15em]
		Bulge+Disk & (\ref{eq:sersic}, \ref{eq:disk})& 13 \\[0.15em]
		Bulge+Bar+Disk & (\ref{eq:sersic}, \ref{eq:sersic}, \ref{eq:disk})& 17\\[0.15em]
		\hline\\[-0.5em]
	\end{tabular}
\end{table}

\subsection{Galaxy fitting procedure}
\label{sec:gal_fit}

\begin{figure*}
\sidecaption
\includegraphics[width=12cm]{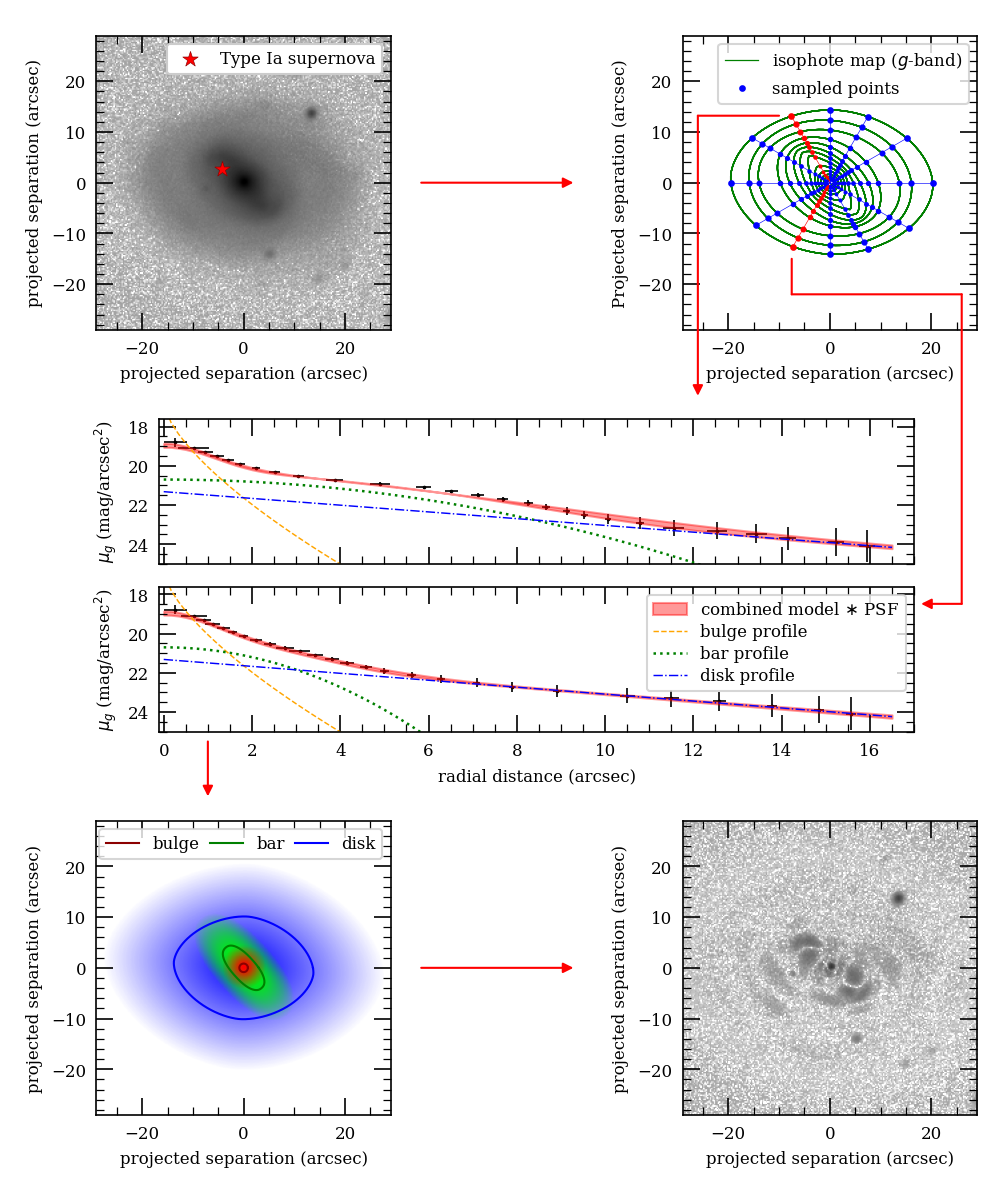}
\caption{Schematic showing the main steps of the galaxy fitting procedure. \textbf{Top-left:} DESI-LS image of a barred host galaxy, containing a ZTF SN Ia. \textbf{Top-right:} The green superellipses are the isophotes of the constructed $g$-band isophote map (every second isophote has been removed for visual clarity). The blue dots mark the sampled points used during the galaxy fitting. \textbf{Middle panels:}  Two example fitted surface brightness profiles are shown. The first panel is located at 120$^\circ$, close to the semi-major-axis of the bar, while the second panel is located at 240$^\circ$, close to the semi-minor-axis of the bar. The black points correspond to the sampled blue dots from the top-right panel, with 3$\sigma$ error bars. The coloured lines (orange, green and red) represent the fitted intrinsic profiles for each component in the `Bulge+Bar+Disk' model. The shaded red region shows the 3$\sigma$ confidence interval for the combined model, convolved with the PSF. \textbf{Bottom-left:} The reconstructed 2D surface brightness model of the galaxy, convolved with the PSF. The three superellipses represent the intrinsic effective radius for each component in the model. The shaded regions are computed by linearly mapping the surface brightness profile to the opacity. The central brightness has an opacity of one, while zero opacity is at the 25 mag/arcsec$^2$ isophote. \textbf{Bottom-right:} Subtraction of model from original image in surface brightness space.}
\label{fig:procedure}
\end{figure*}

Galaxy fitting is the procedure that takes the 1D Sérsic profiles (equations~\ref{eq:sersic} and \ref{eq:disk}) and constructs a 2D surface brightness profile of the galaxy, $\mu(R,\theta)$. In this paper, the effective radius $R_e$, along some angle $\theta$, is given by the superellipse polar equation:
\begin{equation}
    R_e(\theta) = \left(\left|\frac{\cos(\theta-\phi_e)}{a_e}\right|^{c_e} + \left|\frac{\sin(\theta-\phi_e)}{b_e}\right|^{c_e}  \right)^{(-1/c_e)}
	\label{eq:Re}
\end{equation}
where $a_e$ and $b_e$ are the effective radii along the semi-major and semi-minor axis respectively, $\phi_e$ is the position angle of the semi-major axis and $c_e$ is the superellipse index. The same equation also applies to the disk scale length, $h$. This superellipse is taken to be common between both the $g$- and $r$-bands, along with the Sérsic index, while the effective surface brightness $\mu_e$, is allowed to vary between the bands. This assumes that the structure of the galaxy, i.e. the geometric distribution of stars, is the same when observing in either band, but the brightness coming from any particular region will be different in both bands.

The fitting procedure begins with sampling the isophote map along radial lines spaced at $30^\circ$, as seen in the top right of Fig.~\ref{fig:procedure}. The $g$-band isophote map is sampled starting at $0^\circ$, while the $r$-band map is sampled starting at $15^\circ$. This results in the galaxy being sampled along twenty-four radial directions across the $g$ and $r$-bands. The angles are measured anti-clockwise from the positive x-axis (negative Right Ascension), with the point of rotation being the galaxy centre. Since isophotes have varying centres, a correction is required for the sample angle $\theta$, so that the isophotes are sampled along straight lines. This corrected sample angle $\phi$, is found by numerically solving,
\begin{equation}
    d_c \sin(\theta - \beta) - R(\phi)\sin(\phi - \theta) = 0
	\label{eq:stable}
\end{equation}
where $d_c$ and $\beta$ are the distance and relative angle between the galaxy origin and the target isophote centre respectively, and $R(\phi)$ is the sampled superellipse point of the target isophote.

The fitting is performed by the ODRPACK implementation found in SciPy \citep{odr, scipy}. This package performs orthogonal distance regression (ODR), i.e. weighs the fit based on variance in both the x and y-axis. The uncertainties in surface brightness are calculated by taking the mean variance of the pixel values used in the corresponding isophote fit, while the uncertainty in the radial direction is found by error propagation through equation~(\ref{eq:Re}). The 1D galaxy profiles are simultaneously fitted to each radial direction across both $g$ and $r$-bands (middle panels of Fig.~\ref{fig:procedure}), which subsequently constructs a 2D superellipse describing the effective radii/ scale lengths. These superellipses, coupled with the fitted effective surface brightness and Sérsic index, creates a 2D surface brightness profile (bottom left, Fig.~\ref{fig:procedure}).

\subsection{Point spread function}
\label{sec:psf}
The effects of atmospheric seeing on image decomposition are significant and requires careful consideration. The PSF spreads flux away from the galaxy centre, artificially diffuses light into the halo and tends to make elliptical sources more circular (the latter effect requires the PSF correction to be modelled in 2D). In this paper, the PSF is modelled using a 2D circular Moffat function,
\begin{equation}
    \mathrm{PSF}(R) = \frac{\beta-1}{\pi\alpha^2}\left[1 + \left(\frac{R}{\alpha}\right)^2\right]^{-\beta}
	\label{eq:moffat}
\end{equation}
where the power index is chosen to match turbulence theory of our atmosphere, $\beta = 4.765$ \citep{psf}. The core-width $\alpha$, is related to the full width at half maximum (FWHM) by $\textrm{FWHM} = 2\alpha\sqrt{2^{1/\beta}-1}$, where the $g$- and $r$-band PSF FWHM for each sky brick is supplied by DESI-LS. The mean FWHM is 1.51 arcsec in $g$-band and 1.36 arcsec in $r$-band. The PSF is corrected for by numerically convolving Equation~\ref{eq:moffat} with the galaxy model (in flux space) during each iteration of the ODR fitting routine.

\subsection{Initial conditions and bulge truncation}
\label{sec:init}

Continuing the approach used in this paper, the galaxy fitting procedure has been designed to require zero input from the user. Therefore, each galaxy fit requires robust initial conditions that maximise the probability of the ODR fitting routine of converging on the correct solution for any type of galaxy. In addition, the ODR fitting routine is not bounded, so initial conditions must be close to the true solution, to minimise the time searching for a solution.

The single component models (`Elliptical' and `Disk-only') are run first. Having only a single component makes the fitting easy, as there is only one solution. The central surface brightness for both models is initialised to the mean surface brightness of the nine closest pixels to the galaxy centre (defined as the median of all fitted isophotes). The semi-major-axis of the  effective radius/scale length is initialised to the mean semi-major-axis of all isophotes used for the fit. The same is repeated to initialise the semi-minor-axis and the position angle, while the superellipse index is initialised to an ellipse ($c_e=2$). The Sérsic index for the elliptical galaxy fit is initialised to $n=4$ (de Vaucouleurs' profile).

For the two component `bulge+disk' fit, the disk is initialised to the results of the `disk-only' fit, except for the superellipse index, which is reset to two. The bulge central brightness is initialised to the mean surface brightness of the nine closest pixels to the galaxy centre, and the bulge Sérsic index is initialised to one. The superellipse describing the effective radius of the bulge is initialised to the isophote whose corresponding surface brightness is closest to the initialised surface brightness.

\begin{figure}
	\includegraphics[width=\columnwidth]{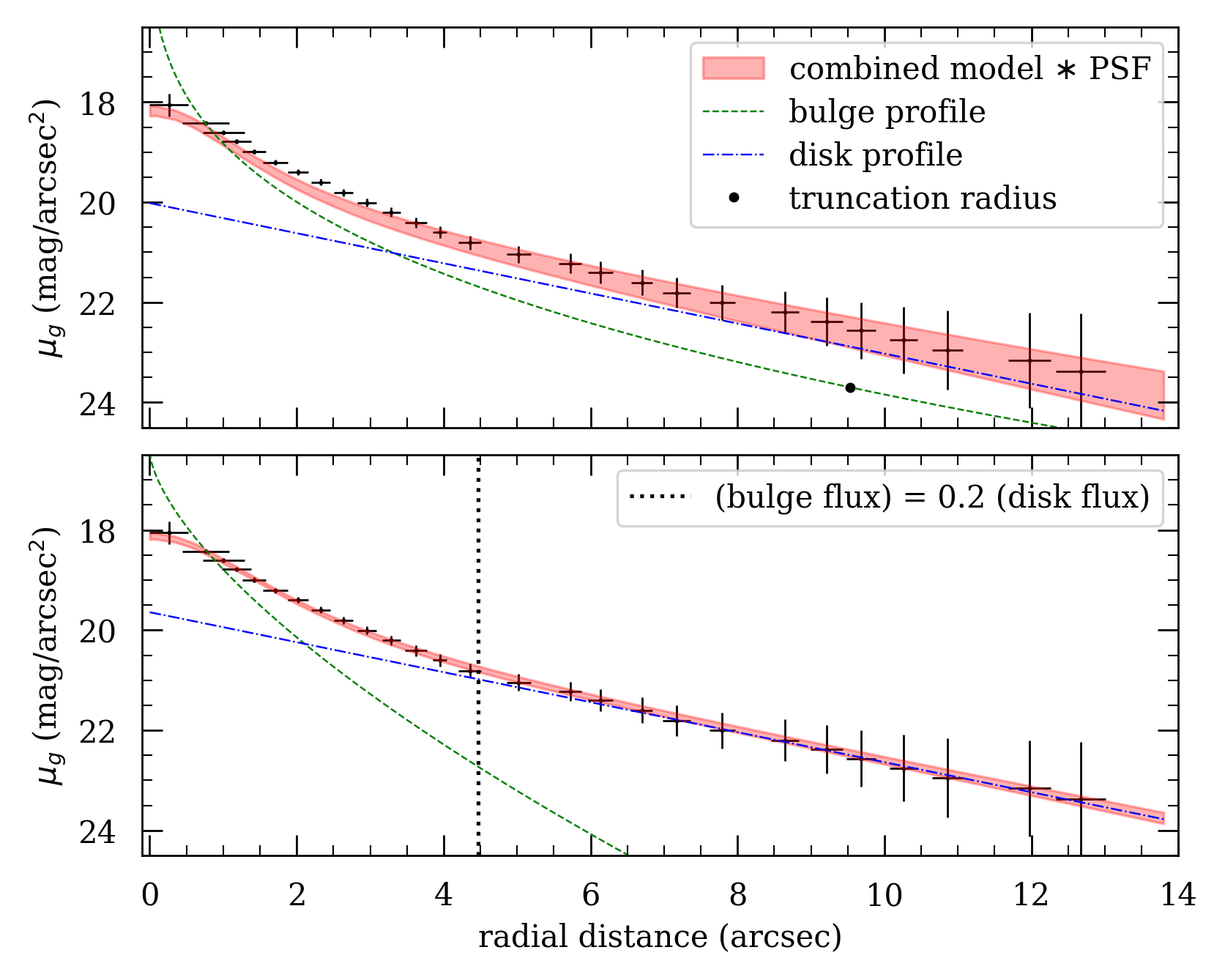}
    \caption {Results of a two-component `bulge+disk' fit to a galaxy. The surface brightness in the \textit{g} band is shown as a function of radial distance from the centre (along an arbitrary angle). The black points are the radially sampled points from the corresponding isophote map for this galaxy, with 3$\sigma$ error bars. The coloured lines (green and red) represent the fitted intrinsic profiles for each component in the `Bulge+Disk' model. The shaded red region shows the 3$\sigma$ confidence interval for the combined model, convolved with the PSF. The top panel has no bulge truncation and converges on a non-physical solution. The bottom panel shows a fit using the same initial conditions, but with bulge truncation active during the fitting procedure. The black dotted line in the bottom panel marks the defined boundary between the bulge and the disk.} 
    \label{fig:trunc}
\end{figure}

For the three component fit, the disk is initialised to the results from the two component `bulge+disk' fit (except for the superellispe index, which is reset to 2). The bulge effective surface brightness and Sérsic index are also initialised from the `bulge+disk' fit. The ellipticity of the bulge effective radius is degenerate with the bar ellipticity, hence for the three component fit, the bulge is defined by a circle rather than a superellipse (reduces the number of free parameters from four to one). The bulge effective radius is initialised to the two-component bulge semi-minor-axis. The bar's effective radius is initialised to the bulge effective radius from the two-component fit, with the superellipse and  Sérsic indices being reset to 3 and 0.5 respectively. The bar's central surface brightness is set to be halfway between that of the bulge and the disk.

An effective approach to further prevent non-physical solutions, is by truncating the bulge profile during the fitting procedure. In this paper, the bulge is truncated at the point where the slope of the bulge profile equals that of the disk. Through differentiation of equations \ref{eq:sersic} and \ref{eq:disk}, the truncation radius is given by
\begin{equation}
    R_{trunc} = R_e \left(\frac{n R_e}{b_n h}\right)^{(n/(1-n))}.
	\label{eq:trunc}
\end{equation}
Due to the singularity at $n=1$, this truncation only occurs if the bulge Sérsic index is greater than 1.1 (since the bar Sérsic index is always less than one, this truncation will not apply). The result of this truncation is that it prevents the ODR fitting routine from using the bulge profile to fit the outer regions of the galaxy, as can be seen in Fig.~\ref{fig:trunc}. Using the same initial conditions for both panels, the bulge truncation prevents the fit from converging on the non-physical solution.

\subsection{Cosmological corrections}
\label{sec:cosmo}
So far in our analysis, the apparent surface brightness ($\mu$) has been used. To convert to absolute surface brightness ($\mu_{\mathrm{abs}}$), two cosmological corrections are required,
\begin{equation}
    \mu_{\mathrm{abs}} = \mu - 10~\mathrm{\log}_{10}(1+z) - \mathrm{K}(z),
	\label{eq:cosmo}
\end{equation}
where $z$ is the cosmological redshift. The first correction is known as cosmological dimming \citep{tolman1930, tolman1934}, and is due to the break-down of the inverse-square law for propagating light in an expanding universe. 

The second correction is known as a K-correction, which corrects for the cosmological redshifting of a galaxy's spectrum due to an expanding Universe, i.e. corrects the $g$- and $r$-photometric bands to the observed galaxy's rest frame. The underlying spectrum of the galaxy impacts the size of the K-correction, so we use different spectral templates for the different galaxy types. The galaxy templates used are the `elliptical' and three of the disk templates (S0, Sb, Sc) from the Superfit SN spectroscopic typing tool \citep{superfit}, with an extra bulge template coming from \cite{kinney}. The bulge template is used for both the bulge and the bar (bars primarily consist of an older population of stars, similar to a galaxy bulge). 

For the `disk-only' galaxy model, the Sc template is used. To choose the most suitable K-correction template for the disk component of the `bulge+disk' and `bulge+bar+disk' models, we make use of the colour difference between the central bulge and the disk, calculated from the fitted galaxy models. The bulge and three disk templates are integrated in the $g$- and $r$-bands to estimate a reference colour difference between the bulge and the disk, $c_{b-d}$, which is defined as the $g-r$ colour of the bulge minus the $g-r$ colour of the disk for each galaxy disk type (S0, Sb, Sc). The fitted colour difference, derived from the galaxy models, is then matched to the closest template colour difference for use in K-corrections. Based on our templates, we define the following ranges for each disk type; S0: $0 < c_{b-d} < 0.17$ mag, Sb: $0.17 \le c_{b-d} < 0.5$ mag and Sc: $0.5 \le c_{b-d} <2$ mag.  It is important to note that these colour ranges are not used to classify the morphology of a galaxy, but are only used to find the closest matching template to be used for K-corrections.

\begin{figure*}
\sidecaption
\includegraphics[width=12cm]{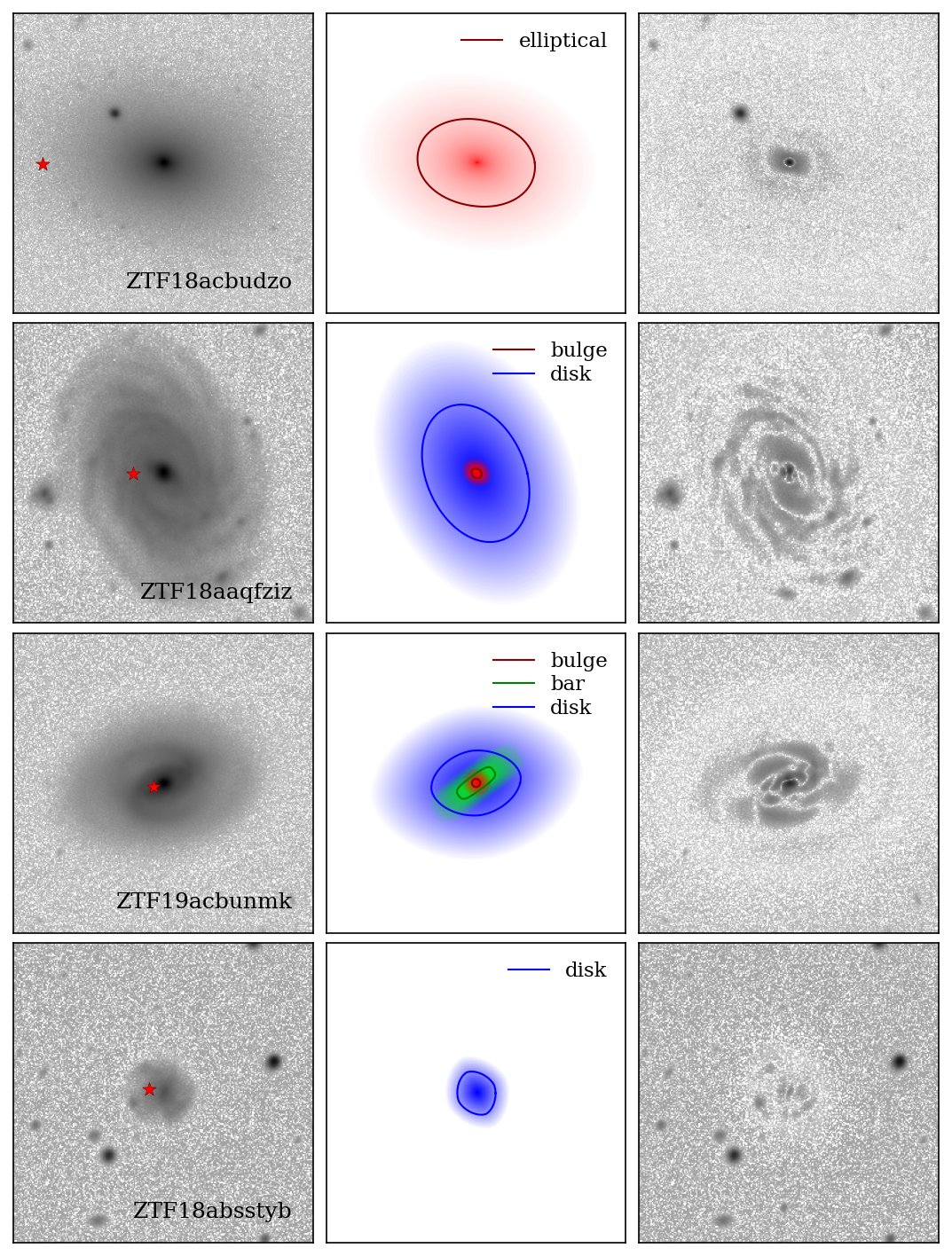}
\caption{Example images of the four galaxy models used in this paper. The left panels are the DESI-LS images in surface brightness space with the location of the SN marked as the green cross, the centre panels are the fitted galaxy models, where the superellipses are the effective radius for each galaxy component. The shaded regions are computed by linearly mapping the surface brightness profile to the opacity. The central brightness has an opacity of one, while zero opacity is at the 25 mag/arcsec$^2$ isophote. The right panels are the residuals, calculated by subtracting the model from the image. The four galaxy models (from top to bottom) are `Elliptical', `Bulge+Disk', `Bulge+Bar+Disk' and `Disk-only'.}
\label{fig:model_img}
\end{figure*}

\subsection{Galaxy morphology identification}
\label{sec:edg_on}
Each galaxy is fitted with all four models listed in Table~\ref{tab:gal_model}. A series of cuts are then applied to determine which galaxy model fits have produced a physically plausible result (see Table~\ref{tab:model_cuts}). The first four cuts are applied to all the models. 

The first general cut is on the residuals of the model and requires that the mean residual is $< 0.02$ mag/arcsec$^2$.  This cut was determined by manually inspecting several model fits, to find the approximate cut-off that is strict enough to only include visually very well fit galaxies, but not too strict as to remove galaxies with spiral arms. The second is a cut on the model disk ellipticity ($1-b/a$) to remove fits with ellipticities of $\geq$ 0.6.  This is because the surface brightness profiles used in this paper describe the face-on profile, and would be inappropriate to use for edge-on systems.  This cut-off value of the ellipticity is estimated from \cite{pastrav2013effects}, who showed that the effects of dust on the derived disk axis-ratio (i.e.~inclination) becomes prominent at an ellipticity of $>$ 0.65. The bulge and disk would also overlap for more edge-on systems, making it difficult to tell where a SN originates from. The `Elliptical' models are also cut at the same ellipticity, to prevent them from fitting edge-on disk systems.

The third cut on all models is based on the value of the superellipse index. The model superellipse cuts are loose bounds, to prevent the inclusion of non-physical shapes, the majority of galaxy fits have values that do not come close to these boundaries. 

\begin{table}
	\centering
	\caption{Parameter cuts applied to model fits. }
	\label{tab:model_cuts}
	\begin{tabular}{l l} 
        \hline\\[-0.5em]
		Model & Parameter cuts\\[0.15em]
        \hline\\[-0.5em]
        All models & Mean residual, $< 0.02$ mag/arcsec$^2$\\[0.15em]
                   &Model ellipticity, $< 0.6$\\[0.15em]
                   &Superellipse index, $1.5 < c_e < 3.5$\\[0.15em]
                   &Model colour, $0 < \mu_{g-r} < 2$\\[0.15em]
		\hline\\[-0.5em]
		Elliptical & Model colour, $\mu_{g-r} > 0.5$ \\[0.15em]
                     &Sérsic index,  $n > 1$\\[0.15em]
        \hline\\[-0.5em]
        Disk-only & Model colour, $\mu_{g-r} < 0.6$\\[0.15em]
        \hline\\[-0.5em]
		Bulge+Disk & Bulge colour, $\mu_{g-r} > 0.5$\\[0.15em]
                   &Bulge colour - Disk colour $> 0$ \\[0.15em]
                   &Central brightness, Bulge $>$ Disk ($r$-band)\\[0.15em]
                   &Bulge/Disk luminosity ratio $< 0.5$ (\textit{r}-band)\\
         \hline\\[-0.5em]
		Bulge+Bar+Disk & Bulge and Disk cuts are the same\\[0.15em]
                       &Bar ellipticity $> 0.3$\\[0.15em]
                       &Bar colour, $0.5 < \mu_{g-r} < 2$\\[0.15em]
                       &Bar semi-major-axis > Bulge radius\\[0.15em]
                       &Bar superellipse index,  $1.5 < c_e < 10$\\[0.15em]
                       &Bar Sérsic index, $n < 0.9$\\[0.15em]
 		\hline\\[-0.5em]
	\end{tabular}
    \tablefoot{The first row contains cuts that apply to all models, subsequent rows contain cuts specific to each model. The colour of the model $\mu_{g-r}$, is calculated by subtracting surface brightness in $r$-band from $g$-band.
    }
\end{table}

The surface brightness colour $\mu_{g-r}$, is calculated by subtracting surface brightness in $r$-band from $g$-band. While each model has its own colour bounds, the overall lower and upper colour boundaries of 0 and 2 mag are chosen to exclude non-physical models, which primarily arise when a model component fails to fit to one of the bands. As will be seen in Section~\ref{sec:Results_scc}, the colour values range from a minimum of 0.2 mag to a maximum of 0.9 mag.

The single component models are given additional colour cuts to prevent the inclusion of unphysical models. The `Elliptical' model has a lower colour cut of $\mu_{g-r} > 0.5$ mag, while the `Disk-only' has an upper colour cut of  $\mu_{g-r} < 0.6$ mag. These values were determined by manually checking the tail-end of the respective colour distributions. Beyond these two colour cut-offs, well-fit `Elliptical' and `Disk-only' models were found to be unphysical. The `Elliptical' colour cut effectively removes the rare case of star-forming elliptical galaxies, which we choose to remove to keep our final sample cleaner. The `Disk-only' galaxies naturally have blue colours, as redder disks are much more likely to have a significant bulge and hence be fit by one of the multi-component models. These cuts removed seven blue `Elliptical' galaxies and five red `Disk-only' galaxies. The overall results and conclusions of this paper are unaffected by the inclusion or exclusion of these galaxies.

In the overlap region between these two models, $0.5 < \mu_{g-r} < 0.6$ mag, a degeneracy is present as the `Elliptical' model is able to mimic a disk by setting its Sérsic index close to unity \citep{diskbulge}. This is not a concern for identifying elliptical galaxies as the `Disk-only' model cannot fit the profile shapes of elliptical galaxies, however `Disk-only' galaxies will be well fit by both the `Disk-only' and `Elliptical' models in this colour region. To resolve these cases, we simply choose the `Disk-only' model. This relies on the assumption that elliptical galaxies generally do not have exponential surface brightness profiles, which is well-supported by the literature \citep[e.g.][]{lange2015galaxy}.

For the multi-component models, the central bulge has the same colour cuts as the elliptical galaxies (similar stellar populations), the bulge must be redder than the disk, and the bulge must have a greater \textit{r}-band central brightness than the disk. The \textit{r}-band bulge-disk luminosity ratio is limited to $<$ 0.5, to remove any potential ambiguity between the `Elliptical' and `Bulge+Disk' models, i.e.~it prevents the `Bulge+Disk' model from setting the disk brightness very low and using only its bulge component to mistakenly fit an elliptical galaxy. This phenomenon is known as a false disk, which is a common pitfall for automated galaxy image decomposition \citep{falsedisk, diskbulge}.  The cut-off value was chosen as the largest bulge-disk ratio found in table 7 of \cite{Graham2008}.

The bar in the `Bulge+Bar+Disk' model requires extra cuts to ensure that it is real and bar shaped. The bar ellipticity is set to be greater than 0.3, so that it is not used to fit ring- or disk-like objects. The bar has the same colour cuts as the bulge, and the semi-major axis of the bar's effective radius must extend beyond the bulge effective radius. The upper bound of the bar superellispe index is set to 10, to enable the bar to take on a more boxy shape, a common feature of bars \citep[][the value of 10 is a conservative upper bound]{sell}. The bar's Sérsic index must be below unity to generate the correct surface brightness profile. However, the cut is lowered to 0.9, to prevent the model from erroneously switching the bar and disk components.

To further aid in the classification of galaxies, recent results from the Galaxy Zoo project \citep{galzoo} are used. \cite{galzoo_desi} have performed 8.67 million morphology measurements of nearby galaxies using the DESI-LS. These were achieved using deep learning models trained on Galaxy Zoo volunteer data. These results overlap with 41\% of the galaxies in the ZTF DR2 sample, providing useful information for distinguishing between the different galaxy models. Four parameters from this study are used: probability of a disk, probability of a bulge, probability of a bar, and probability of the galaxy merging. If the probability of the galaxy merging is $>$ 0.5, the galaxy is discarded.

If following the cuts from Table~\ref{tab:model_cuts} only one model remains, it is used as long as it is consistent with the Galaxy Zoo measurement (if any exist). For example, if our `Elliptical' model is the only model to pass all the cuts and the disk probability from Galaxy Zoo is less than 0.5 then we select this model. If multiple models pass our parameter cuts, the Galaxy Zoo measurements are used to distinguish between them, e.g.~if the `Bulge+Disk' and `Bulge+Bar+Disk' models both pass the cuts, then the bar probability is used to choose the appropriate model (the bar is chosen if bar probability is greater than 0.5). In the case where no Galaxy Zoo measurement is available, and both `Bulge+Disk' and `Bulge+Bar+Disk' models pass, the `Bulge+Bar+Disk' model is chosen. The most likely scenario for both of these models to pass is when a bar does exist, and the `Bulge+Disk' model has absorbed the bar's profile into that of the bulge. If the `Elliptical' model and any of the disk models both pass the cuts, and no Galaxy Zoo measurement is available, then the galaxy is marked as an ambiguous `E-S0'.

\subsection{Galaxy model fitting results}
\label{sec:gal_fit_res}

The results of the galaxy fitting procedure are summarised in Table \ref{tab:gal_fit_cuts}. From the initial 1995 SN Ia host galaxies, 728 (37\%) galaxies have been successfully fitted, and have an identified galaxy type. The Galaxy Zoo measurements, described in Section~\ref{sec:edg_on}, flagged 111 galaxies as merging and hence these are removed before the galaxy fitting begins. From the initial sample of 1884 host galaxies, 25\% (466) fail the residual cut, whereby all four galaxy models were unable to fit the isophote map with satisfactory residuals. A further 19\% are removed, as they have ellipticities greater than 0.6, i.e. are edge-on systems, which cannot be fit using the models employed in this paper, as discussed in Section~\ref{sec:edg_on}. 

From the remaining 1145 galaxies, a further 353 are removed due to the fitted model parameters being unphysical, i.e. the fitted models did not pass all the cuts detailed in Table \ref{tab:model_cuts}. These cuts are applied simultaneously to the fitted models, and generally when a galaxy model fails these cuts, it usually fails a combination of them. Therefore, it is not possible to give a breakdown of how many galaxies fail each of the cuts in Table \ref{tab:model_cuts}. From manually checking the failed models, the main reasons for failure include; highly asymmetric galaxies, merging galaxies (that were not flagged by Galaxy Zoo), very prominent spiral arms/globular clusters/dust lanes, the fitting routine converging on non-physical solutions, a high atmospheric PSF FWHM relative to the galaxy size and intervening Milky Way stars. 

Following the procedure detailed in Section~\ref{sec:edg_on}, the galaxy type (Table \ref{tab:gal_model}) is identified for 728 galaxies, while 64 galaxies were removed due to multiple models passing the cuts, and having no Galaxy Zoo measurement to resolve them.  Fig.~\ref{fig:model_img} shows four example fitted galaxy models, one for each model type.

From the final sample of 728 galaxies, 448 galaxies have a Galaxy Zoo measurement that can be used to benchmark the custom morphology identification used in this paper. The agreement for elliptical galaxies is 92\% while the agreement for galaxies containing disks is 90\%. The reasons for disagreement between our morphologies and those from Galaxy Zoo are predominantly due to the difficulty in distinguishing between elliptical and lenticular galaxies. 

 \begin{table}
	\centering
	\caption{A breakdown of the galaxy model fitting results for SN Ia host galaxies in ZTF Cosmo DR2.}
	\label{tab:gal_fit_cuts}
	\begin{tabular}{l c c c}
        \hline\\[-0.5em]
		Cut & Count& Removed & \% Removed\\[0.15em]
		\hline\\[-0.5em]
        Isophote map fitted & 1995 & -& -\\[0.15em]
        \hline\\[-0.5em]
        Non-merging galaxy & 1884 & 111 & 5.6\\[0.15em]
        Residual cut & 1418 &  466 & 24.7\\[0.15em]
        Face-on galaxy& 1145 & 273 & 19.3\\[0.15em]
        Physical model & 792 & 353 & 30.8\\[0.15em]
        Identified galaxy type & 728& 64 & 8.1\\[0.15em]
		\hline\\[-0.5em]
	\end{tabular}
\end{table}

\subsection{Association of a SN to a galaxy model component}
\label{sec:sn_component}
In the multi-component galaxy models, to determine which model component each SN Ia is associated with, the intrinsic flux ratio at the SN position is used. If at the location of the SN, the flux coming from the bulge or bar is greater than 20\% of the disk flux, then it is classified as a `bulge/bar SN'. The SN is classified as a `disk SN' if the bulge and bar components are $<$ 20\% of the disk flux, and the surface brightness at its location is brighter than 25 mag/arcsec$^2$ in the $r$-band (conventional isophote for the galaxy edge). If the SN is outside this isophote, it is marked as a `halo SN'. This boundary between the bulge/bar and the disk is shown in the bottom panel of Fig.~\ref{fig:trunc}. This value was chosen as it is the approximate flux ratio where the combined surface brightness profile begins to diverge from the pure exponential disk profile. 

The galaxy models used in this paper are fitted to projected images, giving rise to the possibility that the SN lies in front of the associated component and should be marked as a `halo SN'. However, as will be seen in our results, the number of `halo SN' is small and hence this potential misclassification is very unlikely to affect our results. The possibility of misclassification between bulge and disk due to projection effects is also unlikely to be significant. Our final sample does not include any highly inclined systems (Section \ref{sec:edg_on}) that would make this distinction difficult.

\begin{figure*}
    \centering
	\includegraphics[width=17cm]{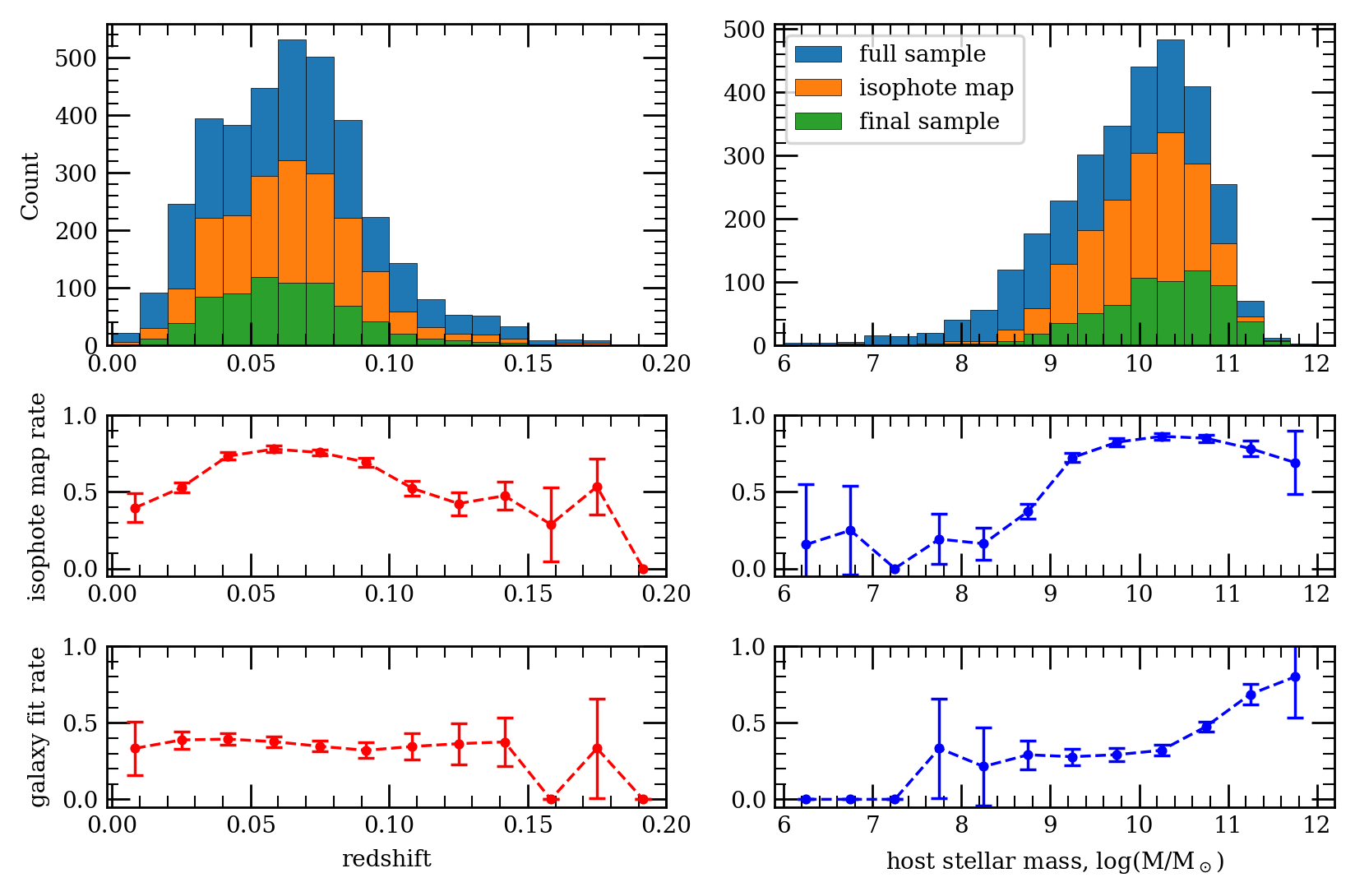}
    \caption {\textbf{Top panels:} Blue histogram shows the full ZTF Cosmo DR2 binned in redshift (left) and host mass (right). Orange histogram shows galaxies with isophote maps that pass all quality cuts. Green histogram shows final sample (successful galaxy fit + galaxy morphology identified). \textbf{Centre panels:} Success rate for generating a successful isophote map, given a galaxy image, as a function of redshift (left) and host mass (right). \textbf{Bottom panels:} Success rate for generating a galaxy model (successful galaxy fit + galaxy morphology identified), given an isophote map, as a function of redshift (left) and host mass (right).} 
    \label{fig:rates}
\end{figure*}

\subsection{SN-component-colour}
\label{sec:scc}
The `SN-component-colour' (SCC$_{g-r}$) is defined as the intrinsic $g - r$ colour (i.e. Milky Way dust, Cosmological and PSF effects removed) of the galaxy component that contains the SN (e.g. Bulge, Bar, Disk or Elliptical). This is found by subtracting the \textit{r}-band from the \textit{g}-band surface brightness of the model, at the location of the SN. The area dependence cancels out, hence the units are given in magnitudes. Since the galaxy shape (effective radius and Sérsic index) is common between both bands during the fitting procedure, each component of the galaxy only has a single colour. Therefore, the SCC$_{g-r}$ captures the average colour of the galaxy component that contains the SN, rather than the colour at the actual SN location, commonly referred to as the local colour.

This colour definition has potential benefits over the local colour for correlating SN Ia properties with their host galaxies. The conditions of the explosion site of a SN Ia can not fully explain its intrinsic properties, as the progenitor system may have orbited several times around its host galaxy before it exploded. Therefore, if a SN is observed to explode within a spiral arm for example (which has a much bluer colour than the disk average), the very blue local galaxy colour of the SN Ia may not better explain its intrinsic properties, as they should be driven more by the environment where the progenitor was born and evolved in. The SCC$_{g-r}$ used in this paper, is an average colour of the entire area that the progenitor system was born and evolved in, and hence could correlate stronger with SNe Ia properties. This approach also removes the issue of having SN light within the galaxy image, as it does not rely on the pixels at the location of the SN. In the case of the `Elliptical' and `Disk-only' models, this colour parameter refers to the global intrinsic colour of the galaxy, as there is only one component in these models. However, for the multi-component galaxies, this colour definition is preferred to the global colour as, for example, it is able to determine the colour of a disk without being biased by the redder flux coming from a central bulge.

\section{Results} 

In Section~\ref{sec:Results_gen}, the results in terms of selection biases against high redshift and low-mass galaxies are provided. Section \ref{sec:final_samp} provides a breakdown of the final sample of host galaxies used to correlate with SN Ia properties. Section~\ref{sec:Results_scc} presents the correlations between the SALT2 light curves parameters (defined in Section~\ref{sec:salt}) and the host galaxy properties of SN component-colour (defined in Section~\ref{sec:scc}) and host stellar mass. In Section~\ref{sec:Results_sb}, the SALT2 parameters are presented in terms of local intrinsic surface brightness, derived from the galaxy models.

\subsection{Redshift and host mass biases}
\label{sec:Results_gen}

As seen in Table \ref{tab:iso_map_cuts}, no successful isophote map was generated for 915 host galaxies, i.e. the isophotes for these galaxies did not pass the quality cuts detailed in Section~\ref{sec:iso_cut}. The majority of these galaxies (693) failed due to the second quality cut that requires at least ten isophotes in both $g$ and $r$ bands. This is a natural consequence of having a fixed image resolution and observational depth. The redshift and host stellar mass distributions of SN Ia host galaxies in the full DR2, those for which a successful isophote map is produced, and those in the final sample for which a galaxy morphology is identified, are shown in the top panels of Fig.~\ref{fig:rates}. The biases in generating isophote maps in terms of redshift and host stellar masses are quantified in the middle row of Fig.~\ref{fig:rates}, which show the rate of generating a successful isophote map given a galaxy image, as a function of redshift and host stellar mass. For redshift, the rate is lower for nearby objects due to the galaxy being too large to fit on a DESI-LS image. The galaxy images are limited to 600 pixels (computational cost is too high for larger images), hence large galaxies at low redshift do not fully fit on the image, and fail the quality cuts. This has the benefit of only including galaxies that are within the Hubble flow, which ensures that the redshift gives an accurate measure of the galaxy distance. The success rate peaks at a redshift of around $z=0.06$ and begins to drop at a redshift of $z=0.1$. No galaxy with a redshift greater than $z=0.18$ possessed a usable isophote map, however these are also very rare events in the DR2 sample. For a fixed pixel scale, the number of pixels per galaxy drops with increasing redshift, hence small galaxies begin to fail to produce isophote maps at high redshifts. For host mass, a similar trend is apparent. The success rate is around 0.9 for higher mass galaxies but begins to drop below a host mass of $\sim10^9$ M$_{\odot}$. The lower the host mass the dimmer it will be, as well as being smaller in angular size, hence the pixel scale and observing depth limitation will create a bias against low mass galaxies from being included in the final sample. This bias is unavoidable due to the limitations of galaxy image decomposition methods such as this one.

The success rate of generating a galaxy model fit that passes the required cuts, given an isophote map, is shown on the bottom panels of Fig.~\ref{fig:rates}. The success rate of generating a galaxy fit as a function of redshift is reasonably constant at $\sim40$\%. There is larger scatter in the highest redshift bins due to small number statistics.  The average success rate for generating a galaxy fit as a function of host stellar mass is $\sim70$\% for the most massive galaxies ($>$10$^{11}$ M$_{\odot}$) and is $\sim40$\% for galaxies with masses of $10^{7.5}$ -- 10$^{11}$ M$_{\odot}$. More massive galaxies have more pixels covering them and are brighter, increasing the successful galaxy model fit rate. Elliptical galaxies are the easiest type of galaxy to fit as they only have a single component and are in general very massive/luminous, hence contributing to the increased success rate for massive galaxies. No galaxy with a mass less than $10^{7.5}$ M$_{\odot}$ is in the final sample. Such galaxies are likely too small/young to have formed pronounced disks, hence none of the galaxy models used in this paper can properly describe these galaxies (e.g.~faint blue galaxies, dwarf irregular galaxies, etc.). This is a notable limitation of our study in that our sample is significantly biased, with an over-representation, compared to the intrinsic rate, of SNe Ia occurring in the most massive galaxies. However, we can still investigate trends and correlations that do not depend on absolute or relative rates of host galaxies. 

\begin{table}
	\centering
	\caption{A breakdown of the final sample of host galaxies.}
	\label{tab:finalsample}
	\begin{tabular}{l c c}
        \hline\\[-0.5em]
        Model & Count & Percentage \\[0.15em]
        \hline\\[-0.5em]
        Full sample & 3628 & - \\[0.15em]
        Final sample & 728 & 20.0 \\[0.15em]
		\hline\\[-0.5em]
		Elliptical & 371 & 51.0\\[0.15em]
        Disk-only & 221 & 30.4\\[0.15em]
		Bulge+Disk & 62 & 8.5\\[0.15em]
		Bulge+Bar+Disk & 74 & 10.1\\[0.15em]
        \hline\\[-0.5em]
       
	\end{tabular}
\end{table}

\subsection{Final sample statistics}
\label{sec:final_samp}

A breakdown of the final sample of host galaxies is presented in Table \ref{tab:finalsample}. Out of our final sample, 51\% are classified as elliptical galaxies. To quantify how biased this value is, we compare this value with the expected rate of SNe Ia in elliptical versus disk galaxies. We use the results from \cite{li2011nearby} who have calculated the rates of SNe Ia as a function of Hubble type, per unit solar mass. We also make use of \cite{kelvin2014galaxy} who have estimated the distribution of mass across each of the galaxy Hubble types. By weighting the SNe Ia rates by this mass distribution, we find that the expected rate of SNe Ia in elliptical galaxies should be approximately $\sim37$\%. Our ratio of elliptical galaxies to disk galaxies of 51\% is higher than the expected rate, indicating a bias towards ellipticals in our final sample. This can be understood by the fact that elliptical galaxies are smooth and featureless, and hence are much easier to fit. Furthermore, as discussed in Section \ref{sec:edg_on}, a fraction of the disk galaxies which are edge-on, were excluded from the final sample, decreasing the total number of disk galaxies. 

Within the disk sample, 62\% have no detected bulge or bar. This number is higher than the true rate due to the physically small nature of bulge/bar systems relative to the disk. Very small bulges and/or bulges at high redshift can become too small and have an insufficient number of pixels to fit a bulge model, resulting in the galaxy being classified as disk only. A very large PSF FWHM can also have this effect, as it can spread a significant amount of bulge flux away from the centre. 

\begin{figure}
	\includegraphics[width=\columnwidth]{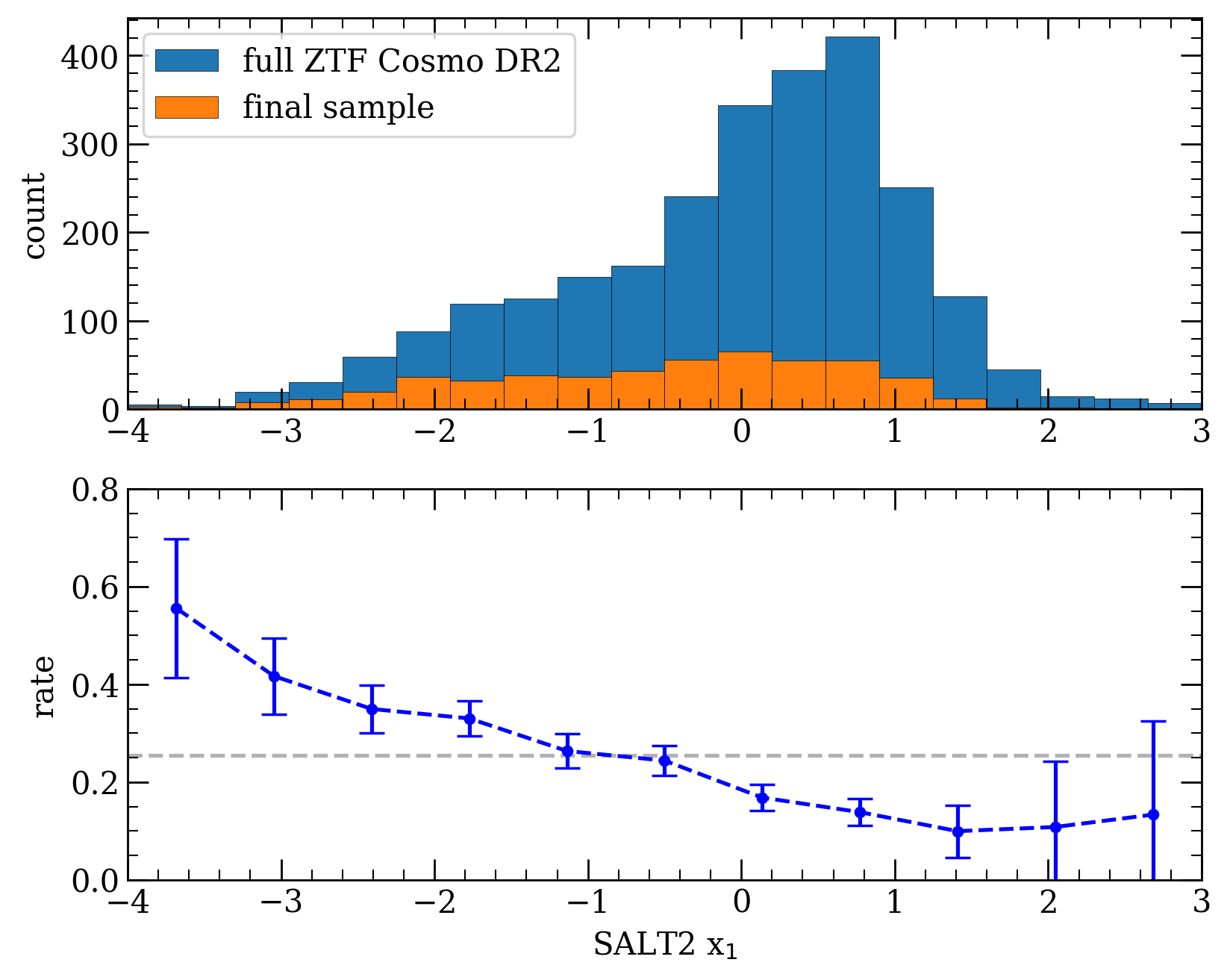}
    \caption {\textbf{Top panel:} Blue histogram shows the ZTF DR2 binned in SALT2 $x_1$, with the light curve quality cuts applied, see Section~\ref{sec:salt}. The orange histogram shows the final sample containing SNe Ia with successfully fitted galaxy models. \textbf{Bottom panel:} Rate of finding a SN Ia with any particular $x_1$ value in the final sample. The dashed grey line shows the average rate.} 
    \label{fig:rates_x1}    
\end{figure}

In Fig.~\ref{fig:rates_x1} in the top panel, we show the SALT2 $x_1$ distribution of the full ZTF DR2 and our final SN Ia sample with a successful galaxy morphological classification (our light curve quality cuts are applied to both samples). In the bottom panel of Fig.~\ref{fig:rates_x1}, we show the rate of a SN Ia with any particular $x_1$ value, entering our final galaxy sample. There is a clear trend, with an excess of SNe Ia with negative $x_1$ values (with respect to the average $\sim25$\%), while there is a deficit of SNe Ia with positive $x_1$ values in the final sample. As will be seen in Section~\ref{sec:Results_scc}, low stretch (negative $x_1$) SNe Ia are predominately found in passive environments, i.e.~elliptical galaxies, while high stretch  (positive $x_1$)  SNe Ia are predominately found in star-forming regions, i.e.~disk galaxies. Therefore, Fig.~\ref{fig:rates_x1} reiterates the result that there is an excess of elliptical galaxies relative to disk galaxies in our final sample. However, there is still broad coverage of the $x_1$ distribution to investigate the correlations between host environment and $x_1$. No apparent bias is seen in the SALT2 $c$ distribution.

 Table \ref{tab:spec_lc_cuts} shows the number of SNe Ia in each morphological component (elliptical, disk, bulge/bar and halo), in our final sample, split based on their spectral properties, and after additionally applying the light-curve quality cuts described in Section~\ref{sec:salt}. The `halo' column is split into SNe Ia found in the haloes of elliptical and disk containing galaxies, i.e., they have local $r$-band surface brightness $\mu_r > 25$ mag/arcsec$^2$. Host galaxies with large outlier SN-component-colour values were manually inspected. This resulted in the removal of nine host galaxies (1.2\%), reducing the final sample size from 728 to 719. The host galaxies removed are those associated with the SNe Ia; ZTF18abugthp, ZTF19abhzewi, ZTF20abffaxl, ZTF18abuykiu, ZTF19aatgznl, ZTF19abquwvx, ZTF19abzkiuv, ZTF20abxidyb and ZTF20abdxuew. By inspection, these galaxies have all converged on a non-physical `Bulge+Disk' or `Bulge+Bar+Disk' decomposition solution.

 \begin{table} 
	\centering
	\caption{Breakdown of the SN Ia morphological component. }
	\label{tab:spec_lc_cuts}
	\begin{tabular}{l c c c c}
        \hline\\[-0.5em]
        Cuts & Elliptical & Disk & Bulge/Bar & Halo(E/D) \\[0.15em]
        \hline\\[-0.5em]
        Full sample & 349  & 275 & 57 & 22/16  \\[0.15em]
        \hline\\[-0.5em]
        Subtype & & &\\[0.15em]
        \hline\\[-0.5em]
         Normal & 214 & 208 & 39 & 15/10  \\[0.15em]
         91T-like & 4 & 11 & 6 & 1/0  \\[0.15em]
         91bg-like & 24 & 0 & 3 & 5/1  \\[0.15em]
         \hline\\[-0.5em]
        Light-curve& & &\\[0.15em]
        \hline\\[-0.5em]
         Normal & 151 & 144 & 24 & 13/8  \\[0.15em]
         91T-like & 2 & 8 & 3 & 0/0  \\[0.15em]
         91bg-like & 11 & 0 & 1 & 3/0  \\[0.15em]
         \hline\\[-0.5em]
	\end{tabular}
    \tablefoot{The `halo' column is split into SNe Ia in the haloes of elliptical (left) and disk containing galaxies (right).}
\end{table}

\subsection{The link between SN-component-colour and host stellar mass}
\label{sec:Results_scc}

\begin{figure}
	\includegraphics[width=\columnwidth]{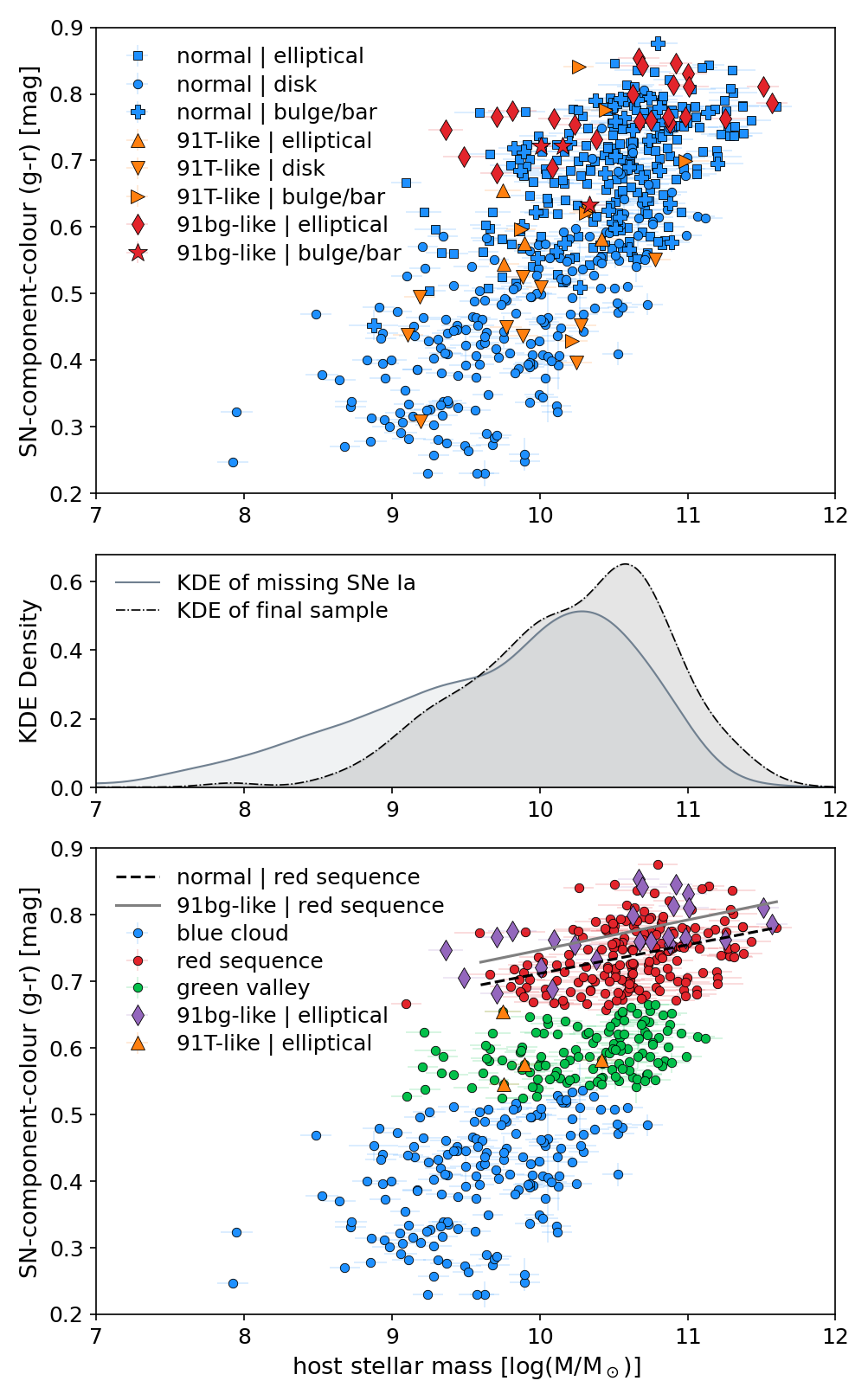}
    \caption {SN-component-colour (\textit{g - r}) versus  host stellar mass. \textbf{Top panel:} SNe Ia separated into the main spectral subtypes, normal SNe Ia (blue), `91bg-like' SNe Ia (red), and `91T-like' SNe Ia (orange). 
    \textbf{Centre panel:} Kernel density plots showing host mass density function of the final sample (dashed line) and the ZTF DR2 SNe Ia that were cut from the final sample (solid line).
    \textbf{Bottom panel:} Normal and `91T-like' SNe Ia grouped into three main clusters: red sequence, green valley, and blue cloud, using a clustering algorithm. The elliptical `91bg-like' (purple diamonds) and elliptical `91T-like' (orange triangles) subpopulations are highlighted. The black-dashed and solid-gray lines are linear regressions for the normal and `91bg-like' subpopulations in the red sequence, respectively.}
    \label{fig:sn_mass}    
\end{figure}

\begin{figure*}
    \centering
	\includegraphics[width=17cm]{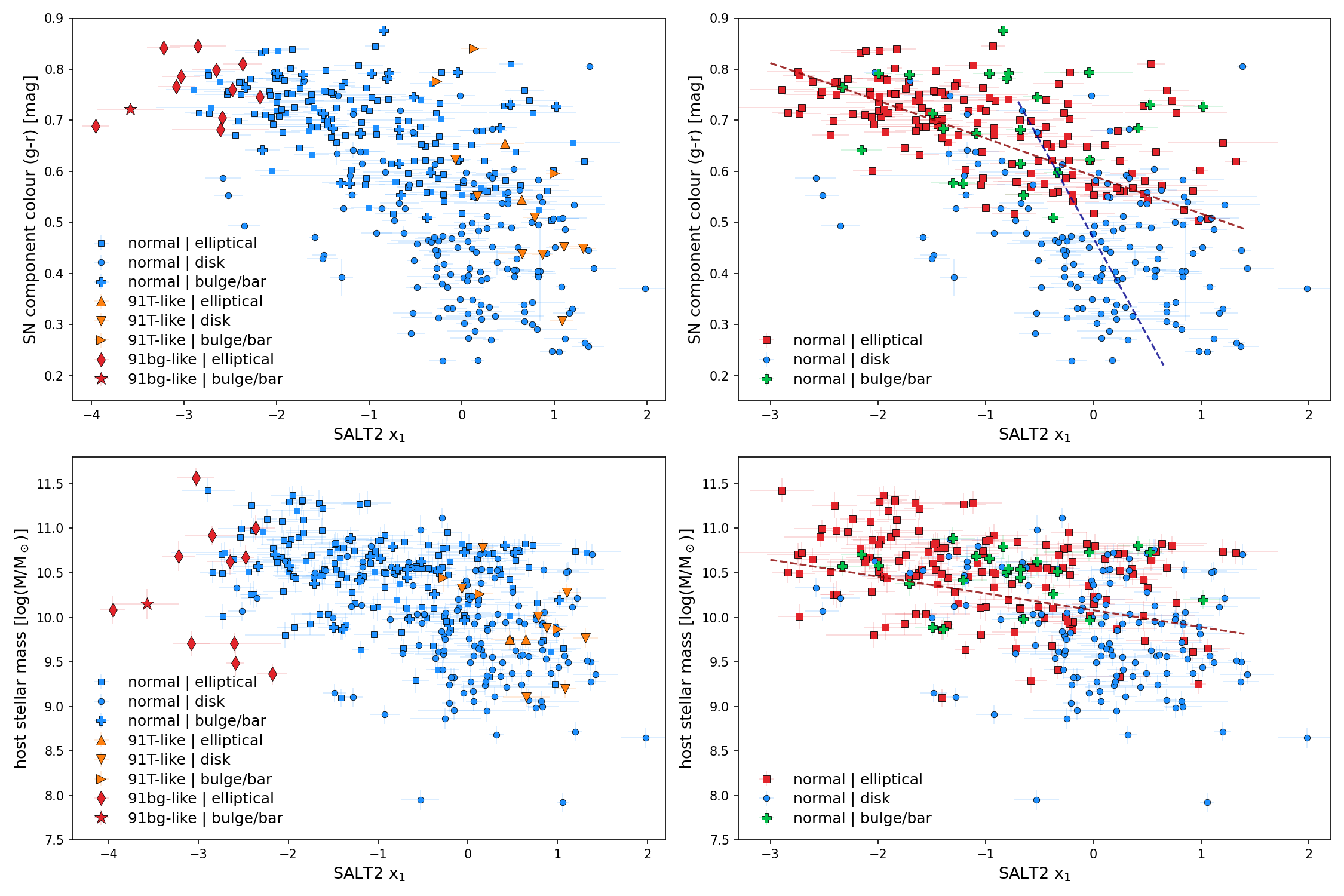}
    \caption {The top panels show SN-component-colour (Section~\ref{sec:scc}) versus $x_1$, while the bottom panels show host stellar mass versus $x_1$. The left panels show the distribution of normal SNe Ia (blue circles), `91bg-like' SNe Ia (red diamonds), and `91T-like' SNe Ia (orange triangles). The right panels contain only the normal SNe Ia, with the associated galaxy component highlighted (Section~\ref{sec:sn_component}). Weighted linear regressions are shown for statistically significant relationships of SN-component-colour/stellar mass and $x_1$ in elliptical galaxies (red dashed lines) and SN-component-colour/stellar mass and $x_1$ in disk galaxies (blue dashed line).} 
    \label{fig:sn_col}    
\end{figure*}

\begin{figure*}
    \centering
	\includegraphics[width=17cm]{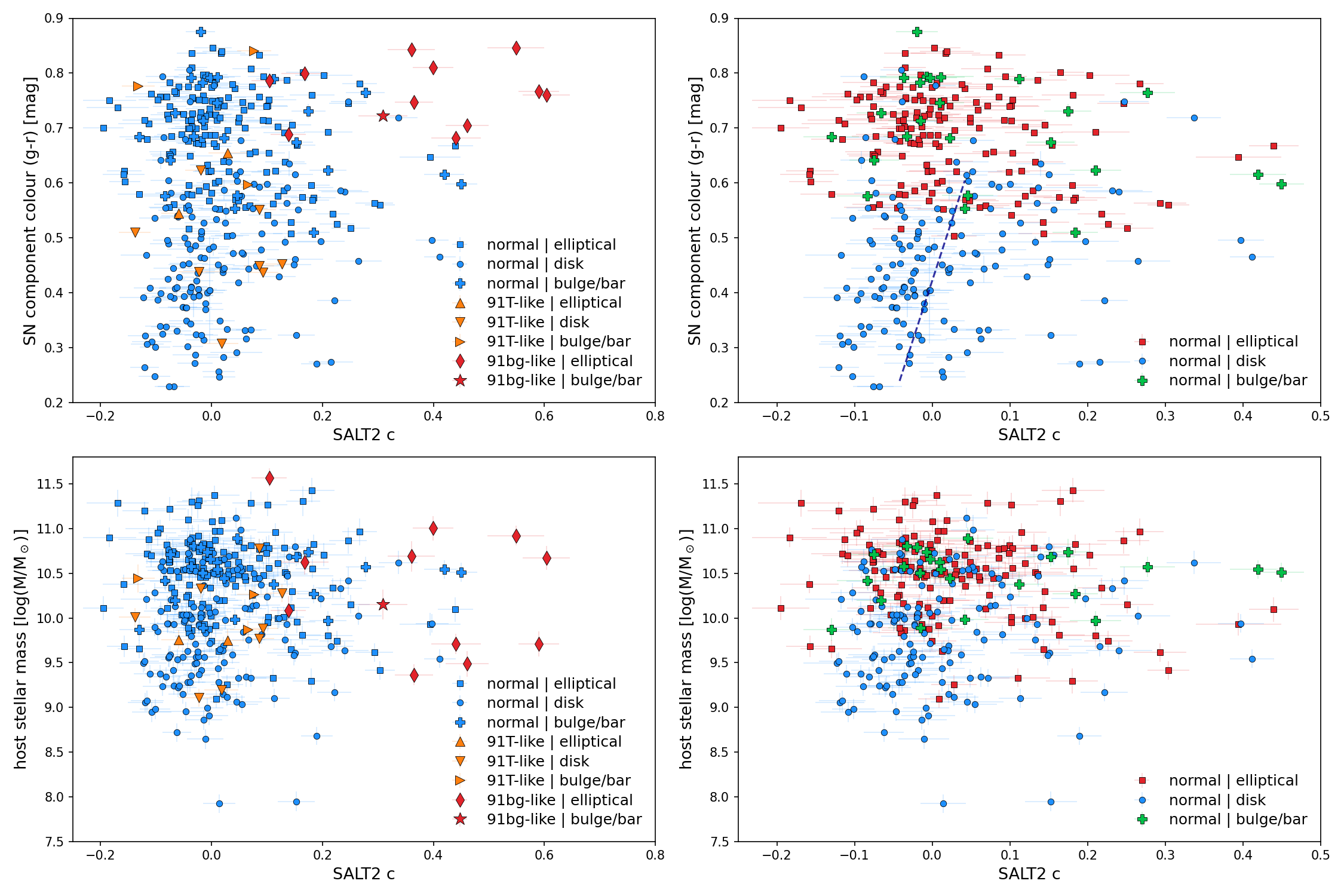}
    \caption {The top panels show SN-component-colour (Section~\ref{sec:scc}) versus $c$, while the bottom panels show host stellar mass versus $c$. The left panels show the distribution of normal SNe Ia (blue circles), `91bg-like' SNe Ia (red diamonds), and `91T-like' SNe Ia (orange triangles). The right panels contain only the normal SNe Ia, with the associated galaxy component highlighted (Section~\ref{sec:sn_component}). A Weighted linear regression (blue dashed line) is shown for SN-component-colour  and $c$ in disk galaxies. } 
    \label{fig:sn_col_c}    
\end{figure*}

In this section, the relation between SN-component-colour (defined in Section~\ref{sec:scc}) and the host stellar mass is investigated. These galaxy properties are plotted in Fig.~\ref{fig:sn_mass} for the final sample with only the subclassification cuts applied. The top panel is divided into the three main SN Ia subtypes based on their spectral classifications (normal SNe Ia, `91bg-like' and `91T-like'). The middle panel shows the density functions of host mass for the final sample of SNe Ia with successful host galaxy modelling, along with the SNe Ia from the ZTF DR2 that were cut from the final sample. As was seen in Fig.~\ref{fig:rates}, we are missing more SNe Ia at lower stellar mass than at higher stellar mass due to biases in our fitting method. The bottom plot is divided into three regions, demonstrating the classical picture of galaxy evolution. This was achieved by using an agglomerative clustering algorithm from scikit-learn \citep{scikit-learn}. The number of clusters was set to three, with the remaining hyperparameters at their default values. The low-mass and blue-coloured galaxies ($\mu_{g-r}\lesssim 0.55$ mag) make up the Blue Cloud \citep{bluecloud}, representing the young galaxies with active star-formation. This population is dominated by disk containing galaxies. The high-mass and red ($\mu_{g-r}\gtrsim0.65$ mag) elliptical galaxies make up the old and non-star forming Red Sequence \citep{blueredgreen}. In between these populations lies the Green Valley \citep{greenvalley}, populated by both elliptical and disk galaxies, representing transitional galaxies with recent star-formation. 

A black dashed line in the bottom panel of Fig.~\ref{fig:sn_mass} shows a linear regression to the normal SN population in the Red Sequence. It can be seen that `91bg-like' events reside above this line. For fixed mass, `91bg-like' events reside in elliptical galaxies that are $0.035\pm0.01$ mag redder (3.5$\sigma$). The slopes of the two (uncertainty-weighted) linear regressions seen in the bottom panel of Fig.~\ref{fig:sn_mass} for normal (5.7 $\sigma$) and `91bg-like' (2.9 $\sigma$) SNe Ia in the red sequence are consistent.  The `91T-like' SNe Ia which reside in the disk, are predominantly found in the Blue Cloud, while `91T-like' events in elliptical galaxies are predominantly contained within the Green Valley. From Table \ref{tab:spec_lc_cuts} it can be seen that within ellipticals galaxies ($z<0.06$), 87\% are normal SNe Ia, 2\% are `91T-like' and 11\% are `91bg-like'. In disk containing galaxies, 92.5\% are normal SNe Ia, 6.1\% are `91T-like' and 1.4\% are `91bg-like'. The `91bg-like' events in disk containing galaxies reside entirely in either the central bulge/bar or in the halo.

In Fig.~\ref{fig:sn_col}, the relationship between SN Ia light-curve stretch (SALT2 $x_1$) and host galaxy SN-component-colour and stellar mass is investigated. The left panels show these colour coded by the SN Ia spectral classification (normal, 91bg-like, 91T-like) and the right panels, show the normal SNe Ia only, separated by the associated galaxy component. This plot, as well as subsequent plots, have the subclassification and light-curve cuts applied. We identify a high significance linear relation (16.8$\sigma$) between SN-component-colour and SALT2 $x_1$ for the normal SNe Ia occurring in the elliptical galaxy population (top right panel of Fig.~\ref{fig:sn_col}). SNe Ia with narrower light curves (more negative $x_1$) occur in redder elliptical galaxies than those with broader light curves (more positive $x_1$).  The Pearson correlation coefficient for this relation is $-$0.61, giving a R$^2$ value of 0.37, implying that up to 37\% of the variation seen in the $x_1$ values of SNe Ia located in elliptical galaxies, is explained by the intrinsic colour of their host environment. The bulge/bar subpopulation has too few SNe Ia for meaningful statistics, but they can be seen to follow a similar trend as the elliptical galaxies. This is not surprising, since they are also expected to be passive environments with older stars than those seen in galaxy disks. Using a Kolmogorov-Smirnov test (two-sided), the probability that the bulge/bar SN Ia $x_1$ distribution is different from the distribution seen in elliptical galaxies is only 1.2$\sigma$, while the probability that they are different from the disk SNe Ia is 4.6$\sigma$. 

A relation between SN-component-colour and SALT2 $x_1$ for the normal SNe Ia occurring in galaxy disks is also identified. This correlation is much steeper than the one seen in elliptical galaxies, as well as much weaker, however it is still significant (5.1$\sigma$). With a Pearson correlation coefficient of $-0.39$, the disk colour explains approximately 15\% of the variation seen in $x_1$ values of SNe Ia occurring in galaxy disks. The 91T-like SNe Ia that occur in disk galaxies appear to follow the same SN-component-colour and $x_1$ relation as the normal SN Ia population. In the bottom right panel of Fig.~\ref{fig:sn_col}, a trend is also seen between  host stellar mass and $x_1$ in the elliptical population.  However, this trend is much less significant (3.3$\sigma$, R$^2$ = 0.13) than that between SN-component-colour and $x_1$ (16.8$\sigma$, R$^2$ = 0.37). 

In Fig.~\ref{fig:sn_col_c}, it is apparent that SN Ia light-curve colour (SALT2 $c$) does not correlate as strongly with host galaxy properties as  $x_1$ does. As has been seen before, the SALT2 colours of 91bg-like SNe are redder, on average, than normal and 91T-like events, and as was seen in Fig.~\ref{fig:sn_mass}, they occur among the reddest elliptical and bulge/bar environments.   In the right panels, the only apparent trend (2.96$\sigma$, R$^2$ = 0.07) is between SN-component-colour and SALT2 $c$, for normal SNe Ia in disks (with the caveat that  outlier points with galaxy colours of $>0.65$ mag are excluded. Including these points drops the significance to 2.5$\sigma$).  Using a Kolmogorov-Smirnov test (two-sided), the probability that the SN Ia colour distribution seen in the elliptical galaxies is different from those in the disk, is only 1.9$\sigma$ (with elliptical SNe Ia having slightly redder colours). No significant trends between host stellar mass and colour are seen in the bottom row of Fig.~\ref{fig:sn_col_c}.

\subsection{Light curve properties and local surface brightness}
\label{sec:Results_sb}

\begin{figure*}
    \centering
	\includegraphics[width=17cm]{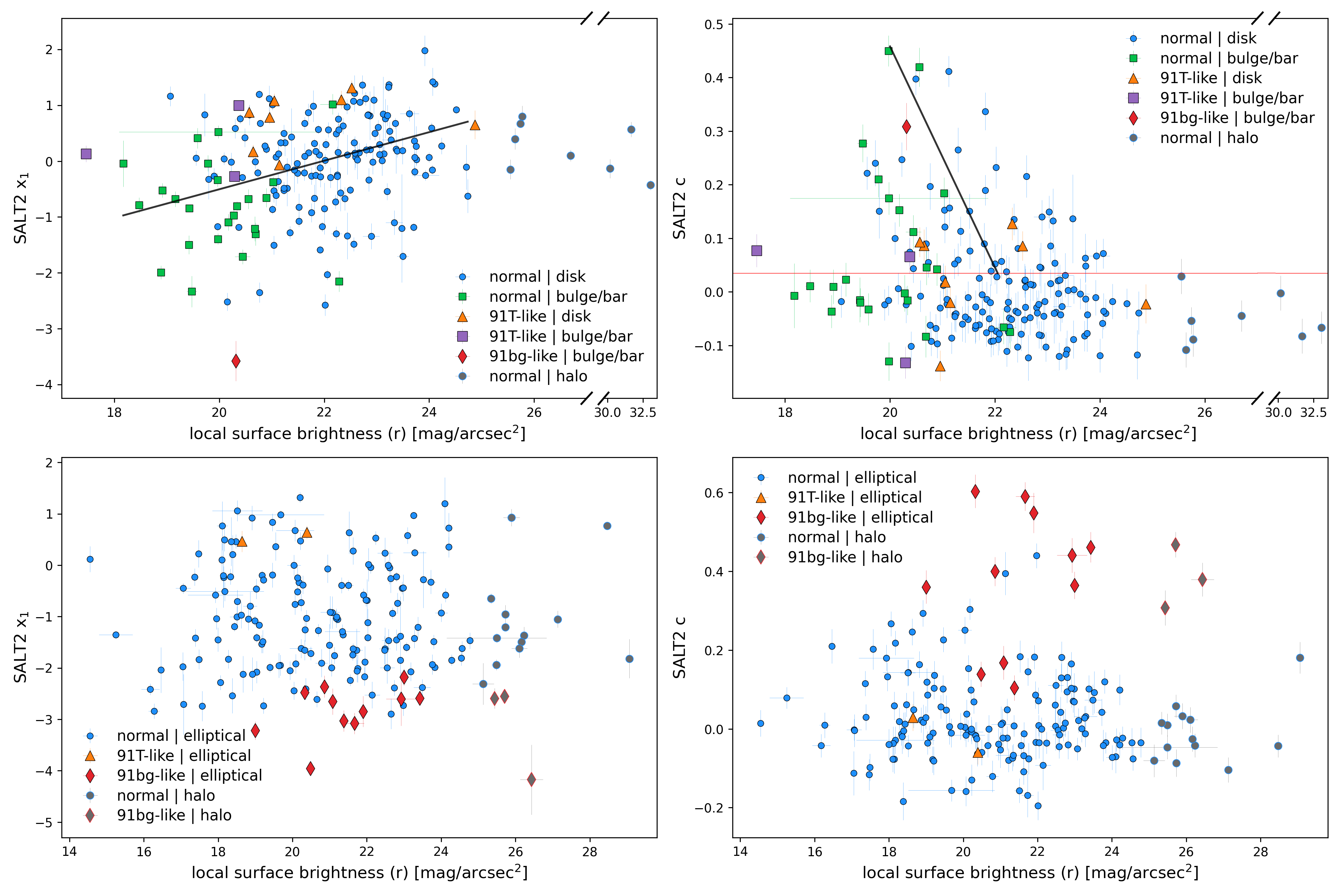}
    \caption {Local $r$-band surface brightness (derived from model) versus SALT2 parameters ($x_1$ on the left and $c$ on the right), highlighted by SN classification and the associated galaxy component. The top panels contain SNe Ia located in galaxies containing a disk, while the bottom panels contain SNe Ia located in elliptical galaxies. Black dashed lines indicated significant linear correlations. A horizontal red line is plotted at $c=0.035$. See Section~\ref{sec:dust} for discussion.} 
    \label{fig:sn_sb}    
\end{figure*}

In Fig.~\ref{fig:sn_sb}, the SALT2 light-curve parameters of $x_1$ (left panels) and \textit{c} (right panels) are plotted against the local $r$-band surface brightness (intrinsic surface brightness of the galaxy model at the position of the SN).  The top panels show SNe Ia occurring in `disk-only', `bulge+disk', `bulge+bar+disk' galaxies and the bottom panels show the SNe Ia that occur in elliptical galaxies. Their membership of a disk, bulge/bar, elliptical, or halo is indicated by the symbols and colours. In the top panels, as expected, the SNe Ia in bulges or bars have, on average, the brightest local surface brightness, followed by SNe Ia in disks, and the SNe Ia in the halo are the regions of lowest surface brightness. In the top-left panel, a significant linear correlation (6.1$\sigma$, R$^2$ = 0.09) is seen between $x_1$ and local $r$-band surface brightness, for the normal SN Ia population originating from galaxies containing disks (disk, bulge or bar). The normal SNe Ia residing in the halo do not appear to follow the same trend and have a mean stretch of $x_1 = 0.23 \pm 0.16$. This stretch value is more consistent with the mean stretch of disk SNe Ia $x_1 = 0.02 \pm 0.07$, than with SNe Ia found in the bulge/bar $x_1 = -0.81 \pm 0.17$.

In the bottom left panel of Fig.~\ref{fig:sn_sb}, no correlation is seen between SALT2 $x_1$ and local $r$-band surface brightness for the elliptical galaxy population of SNe Ia. However, the interpretation of surface brightness for elliptical galaxies is more difficult than for disk  galaxies due to the difficulty in determining where SNe Ia occur within the 3D elliptical galaxies from a projected image. SNe Ia located near the centre of elliptical galaxies (and hence have a bright surface brightness), may in fact lie in front of the galaxy, making the true surface brightness much dimmer. This projected surface brightness problem does not affect disk SNe Ia to the same extent, as the face-on disks used in this study can be considered as 2D objects, and hence the measured surface brightness relates to the true stellar density at the explosion site. The 91bg-like population shown in the bottom left panel of Fig.~\ref{fig:sn_sb} does not appear to reside in brighter or fainter surface brightness components than the normal SNe Ia populations. 

In the top-right panel of Fig.~\ref{fig:sn_sb}, local $r$-band surface brightness is plotted against SALT2 $c$, for the SN Ia population located in galaxies containing disks, highlighted by the SN classification and the associated galaxy component (disk, bulge/bar or halo). A horizontal red line has been plotted at SALT2 $c = 0.035$ (the choice of this value is arbitrary). Below this line, normal SNe Ia are present across all surface brightness values, while above this line,  the normal SNe Ia are localised between $ 19 < \mu_r < 24.5$ mag/arcsec$^2$, and have a significant linear trend (3.3$\sigma$), whereby redder SNe Ia reside at brighter surface brightness. The normal SNe Ia residing in the halo, have a mean SN colour of  $c =-0.052 \pm 0.015$, with all events having  SALT2 $c < 0.035$. In the brightest regions around the galaxy centres (where bulge/bars are located), the apparent cluster of normal SNe Ia ($-0.05 < c < 0.035$ and $\mu_r<20.8$ mag/arcsec$^2$), group tightly at  $c = -0.011 \pm 0.004$.  

In the bottom-right panel of Fig.~\ref{fig:sn_sb}, no significant correlation between SALT2 \textit{c} and local $r$-band surface brightness is seen for normal SNe Ia in elliptical galaxies. However, SNe Ia with  $c > 0.035$ do populate the bright surface brightness values, which are missing from the top-right panel. The `91bg-like' SNe Ia (red diamonds) appear to be absent from the brightest regions ($\mu_r \lesssim 19$ mag/arcsec$^2$) of elliptical galaxies. This is likely an observational bias, as `91bg-like' SNe Ia  are generally less luminous, and therefore, are more difficult to detect near the bright centres of elliptical galaxies.

\section{Discussion}

\subsection{Galaxy morphology and SN properties}

Many studies have shown empirical correlations between SN Ia stretch and host galaxy properties, whereby higher stretch SNe Ia (and hence more luminous SNe Ia) are preferentially found in less massive star-forming environments, while low stretch SNe Ia are predominately found in more massive passive galaxies \citep{hamuy1996, sullivan2003, kim2019, salt_morph, nicolas2021, garnavich2023connecting}. We identify a similar trend between $x_1$ and host stellar mass for the much larger sample of ZTF DR2 SNe Ia studied here (bottom panels of Fig.~\ref{fig:sn_col}), which is also seen in the ZTF companion paper, \customcitecolor{blue}{Ginolin2024a}. 

We have also investigated in a large sample for the first time, the relation between the distribution of SNe Ia in different galaxy components (elliptical, bulge/bar, disk) and SALT2 parameters. It can be clearly seen in the top panels of Fig.~\ref{fig:sn_col} that SNe Ia in elliptical galaxies have a different $x_1$ distribution than those in disk galaxies. SNe Ia located in the disk have a mean $x_1$ of $0.12 \pm 0.06$, while the ellipticals have lower $x_1$ SNe Ia, with a mean $x_1$ of $-1.04 \pm 0.8$. The SNe Ia in elliptical galaxies overlap with the $x_1$ values of disks in the region of $x_1 > -1$, however, the elliptical SNe Ia  account for $\sim 75$\% of the SNe Ia  with  $x_1 < -1$. The few SNe Ia in disks (17\%) with low stretch values ($x_1<-1$) may originate above or below the disk (e.g.~in globular clusters), where the stellar population is much older \citep{barkhudaryan2023}. 

The right panels of Fig.~\ref{fig:sn_col} show that the SNe Ia identified to reside in a bulge/bar have $x_1$ and SN-component-colour values that match the elliptical population. Using a two-sided Kolmogorov-Smirnov (KS) test, there is only a 1.2$\sigma$ level probability of the bulge/bar $x_1$ distribution being different from that of the elliptical $x_1$ distribution.  This supports the idea that the bulge/bar of disk containing galaxies have an older population of stars, similar to those found in elliptical galaxies.

In a ZTF companion paper, \customcitecolor{blue}{Ginolin2024a} investigate the standardisation procedure used in cosmology, focusing on the environmental dependencies of SN Ia stretch and luminosity of the ZTF DR2 sample. They find a significant (14.3 $\sigma$) non-linearity between standardised Hubble residuals and $x_1$, quantified by a `broken-$\alpha$’ model (where $\alpha$ refers to the fitted variable used to standardise $x_1$). They report that the best-fit value to separate the low- and high-stretch modes is at $x_1$ of $-0.49 \pm0.06$. In another ZTF companion paper, \customcitecolor{blue}{Deckers2024} investigate the secondary maximum seen in the light-curves of SNe Ia in the ZTF DR2 sample. They report a non-linearity between the $r$-band integrated flux under the secondary maximum and $x_1$, with their best fit `broken-line’ splitting the low and high-stretch modes at $x_1^0=-0.6\pm0.2$, consistent within the uncertainties with that of \customcitecolor{blue}{Ginolin2024a}. Different versions of this `broken-$\alpha$’ model have been previously seen  \citep{burns2018carnegie,garnavich2023connecting}.  

It is difficult to estimate the point between potential high- and low-stretch dominated regimes for our sample. However, we can attempt this using the intersection point of the two linear regressions between SN component colour and $x_1$ seen in the top-right panel of Fig.~\ref{fig:sn_col}. This intersection point is at a value of $x_1=-0.39\pm0.06$. This $x_1$ value is less negative but consistent with both \customcitecolor{blue}{Ginolin2024a} and \customcitecolor{blue}{Deckers2024}. However, we reiterate that it is difficult to determine whether this intersection is an accurate measure.

\subsection{SNe Ia in elliptical galaxies}
This is the first study that demonstrates that significant linear correlations exist in both elliptical and disk SNe Ia between SN-component-colour/host stellar mass and the SALT2 parameter $x_1$. We have identified a 16.8$\sigma$ correlation between SN-component-colour and $x_1$ for SNe Ia occurring in elliptical galaxies. The correlation has a R$^2$ of 0.37, which points to a fundamental link between SNe Ia $x_1$ and the colour of the galaxy component (bulge, bar, or disk) that the SN resides in. A lower significance correlation (3.3$\sigma$,  R$^2$=0.13) between host stellar mass and $x_1$ is found for SNe Ia in elliptical galaxies, suggesting the SN-component colour is the preferred environmental tracer. A similar result was seen by \cite{kelsey2021}, who found a larger galaxy colour step compared to the host mass step in the Hubble Residuals. 

These trends that are seen when SNe Ia in elliptical galaxies are separated from SNe Ia in disks, as well as the fact that there appears to be a distinct population of SNe Ia in the reddest elliptical galaxies that have significantly smaller $x_1$ values than SNe Ia in the disks of galaxies, requires a physical explanation. To attempt to reach such an explanation, the driving factors behind these correlated parameters must be first identified.  The driving force in the colour of elliptical galaxies is commonly believed to be a function of both stellar age and metallicity. While the relative strength of these driving factors depend on the colour being investigated, \cite{chang2006colours} argue that metallicity is the primary driver of galaxy colours (relative to luminosity-weighted age), but age has the most significant effect on optical colours (such as $g$ and $r$-bands). 

There is a well known relation between galaxy colour and magnitude. The driving force in the colour of elliptical galaxies is commonly believed to be a function of both stellar age and metallicity \citep{thomas2010_ellipticals,Maiolino2019aarv_metallicity}, with age and metallicity both increasing for higher mass galaxies. \cite{thomas2010_ellipticals} provided scaling relation between age/metallicity and dynamical mass for early-type galaxies. However, these scaling relations can not be directly applied since we have host stellar masses and not velocity dispersion measurements that could be used to estimate the dynamical mass of the galaxies. 

Studies of the delay-time distributions (DTD) of SNe Ia have suggested that they are best fit by two-component models (`prompt' and `delayed'), such as \cite{mannucci_rates}. A comparison between SN Ia DTD and age and $\alpha$-element abundances in early-type/elliptical galaxies has suggested these distributions are best fit by a single power-law DTD model \citep{walcher_dtd}, which could imply a single progenitor origin for SNe Ia in elliptical galaxies. We have identified a strong trend between SN component colour of the galaxy and SALT2 $x_1$, for `normal' SNe Ia in galaxies we have classified as elliptical. In our SN Ia sample, nearly all `91bg-like' events are found in elliptical galaxies (there is just one in a bulge/bar location) and within these elliptical galaxies, they occur in the reddest galaxies. The `91bg-like' SNe Ia follow the correlation between SN component colour and SALT2 $x_1$ seen for the `normal' events in elliptical galaxies, suggesting that, from this aspect at least, their progenitor origin may be similar. For a sample of SNe Ia in 29 early-type galaxies, a strong correlation between SN peak \textit{V}-band magnitude and host age (measured by the strength of typical absorption lines in old stellar populations) was identified \citep{Gallagher2008}. `91bg-like' SNe Ia have also been suggested to come from progenitors with delay times of $>$ 6 Gyr \citep{Panther2019}. 

\cite{Shen2017} investigated the relation between SN Ia decline rate and stellar age using binary population synthesis calculations for sub-Chandrasekhar mass double-detonation systems. They found that the fastest declining SNe Ia (equivalent to the most negative SALT2 $x_1$) are associated with the oldest populations. This relation is driven by the fact that binaries with higher mass primaries interact at younger ages compared to systems with less massive white dwarf primaries. Higher metallicity SN Ia progenitors can result in more neutron capture, resulting in more stable Fe-group isotopes at the expense of radioactive Fe-group elements resulting in a fainter SN Ia, but this effect is expected to be small. Therefore, it is not expected that metallicity plays a major role in $x_1$ diversity. This is also seen in studies of galaxy-based forward models investigating the impact of age on SN Ia luminosities, where variations in age are found to be consistent with the SN luminosity variations observed \citep{Wiseman2022,Wiseman2023}. \cite{kim2024_SNIa_early} investigated a population of SNe Ia in elliptical galaxies to attempt to break the degeneracy between age and metallicity in their birth environments, but no clear trends were seen. However,  \cite{Hayden2013} investigated the fundamental metallicity relation for a sample of star-forming hosts and concluded that the correlation between Hubble residuals and host stellar mass is most likely driven by metallicity, highlighting the somewhat conflicting results in the literature.

By linking these studies with our identified trend between $x_1$ and SN component colour, where the SNe Ia with more negative $x_1$ values occur in redder environments, it suggests that if the SN Ia population in elliptical galaxies is due to double degenerate white-dwarf systems (and it appears hard to explain in the context of a Chandrasekhar-mass explosion in these very old systems), then it suggests that the stellar age of the environment is more important than metallicity for the elliptical galaxies hosting SNe Ia.

\cite{briday2022} investigated the best environmental tracers to explain the `mass step' correction applied in SN Ia cosmology. They concluded that local specific star formation rate from H$\alpha$ emission and global stellar mass are better tracers of this galaxy environment step function than host morphology and other tracers such as local mass or local colour. For their morphologies, they used two different methods, for the SN Factory data, the ratio of the radii containing 50 and 90\% of the Petrosian flux in the \textit{r} band \citep{Kauffmann2003} supplemented with literature events with a visual morphological classification of \cite{Pruzhinskaya2020}.

\subsection{SNe Ia in galaxies containing disks}
A 5.1$\sigma$ correlation exists between SN component colour and $x_1$ for SNe Ia occurring in disks (top-right of Fig.~\ref{fig:sn_col}). However, with a R$^2$ of only 0.15, this trend is not as strong as that seen in the elliptical population. It is clear that for the actively star-forming disk galaxies, with a SN-component-colour below $\sim0.55$ mag, there is no trend. Normal SNe Ia originating from actively star-forming environments (e.g.~very blue disks)  typically have $-1<x_1<2$. Disk environments, with $g-r$ SN-component-colour of $>0.55$ mag, start to overlap with the colours of elliptical galaxies (i.e.~in the `Green Valley'). These disks may have quenched star formation (hence redder colours), and therefore, resemble older environments that typically contain lower stretch SNe Ia. Therefore, these `Green Valley' SNe Ia are likely driving the observed correlation between SN component colour and $x_1$.

From Fig.~\ref{fig:sn_col_c}, it can be seen that SN Ia colour shows little correlation with galaxy colour or host mass in galaxies containing disks, which is in agreement with the literature \citep[e.g~][]{sullivan2010, uddin2017influence}. From the top-right panel, it can be seen that there is a weak trend (2.96$\sigma$) in the SNe Ia located in disks, where redder disks tend to host redder SNe Ia. However, with such a large scatter, this result is better interpreted as a lack of red SNe Ia in very blue disks. A similar result was also reported in \citep{sullivan2010, kelsey2023concerning}. The colours of star-forming disk galaxies are generally much harder to relate to metallicity or age, as many star-formation bursts can occur during a disk's lifetime, complicating the age and chemical composition each time. Colours of elliptical galaxies likely correlate better than the colours of disks, as the period of time post star-formation is much longer relative to the timescales between star-formation bursts.

In the top left panel of Fig.~\ref{fig:sn_sb}, a significant linear correlation (6.1$\sigma$) is seen between $x_1$ and the local $r$-band surface brightness in galaxies with disks (`disk-only', 'bulge+disk', `bulge+bar+disk' models). We can deduce that this must be an environmental dependence with the SN Ia progenitors, as a purely observational explanation should also produce a correlation for the SNe Ia in elliptical galaxies, which is not seen (bottom left panel of Fig.~\ref{fig:sn_sb}). In addition, we can rule out any observational bias on the SN Ia magnitude as an explanation, as the trend is in the opposite direction. Lower stretch SNe Ia are generally less luminous \citep{phillips1993absolute}, and hence should be more difficult to observe in brighter regions. However, the trend shows the opposite, with lower stretch SNe Ia being more abundant near the galaxy centre. 

This surface brightness correlation is apparent for disk systems but not for elliptical galaxies and can be explained by considering the distinction between SNe Ia explosion location and progenitor birthplace, as the local host environment at the birthplace should correlate stronger with the intrinsic properties of the SN Ia, rather than the explosion site. For disk galaxies, stars generally orbit in circular orbits within the plane of the disk, resulting in the stellar density (i.e. surface brightness) following the gravitational potential of the galaxy. Therefore, the SN Ia progenitors in disk systems, will likely be born, evolve and explode all at approximately the same surface brightness value, i.e. on the same isophote. In contrast, the stellar orbits within elliptical galaxies are much more chaotic and are generally not circular. Therefore, the measured surface brightness at the SN Ia explosion site in elliptical galaxies, has no correlation with the local surface brightness at the time of progenitor birth or during the progenitor evolution.

Disk galaxies are generally believed to grow inside-out, whereby the oldest population are closest to the centre and active star-formation of young stars occurs further out in the plane of the disk \citep{delgado2015}. Hence, surface brightness, which is directly related to the galactic separation and stellar density, is correlated with both age and metallicity in galaxy disks, where age and metallicity increase towards the centre \citep{macarthur2004structure, sanchez2014}. The observed trend of lower stretch SNe Ia being more abundant closer to the galaxy centre, implies that lower stretch SNe Ia are associated with older and more metal rich environments. This is the same result as that from the SN component colour in Fig.~\ref{fig:sn_col}. Lower stretch SNe Ia reside in redder colours, which is known to also correlate with older and more metal-rich environments.

\subsection{The impact of dust extinction}
\label{sec:dust}
The top-right panel of Fig.~\ref{fig:sn_sb} shows an interesting correlation between SALT2 $c$ and $r$-band surface brightness for SNe Ia located in disk containing galaxies. The 3.3$\sigma$ linear trend, seen in the normal population with SALT2 $c > 0.035$, could be due to dust extinction, as dust content is known to increase towards the centre of disk galaxies  \citep[i.e.~follows surface brightness;][]{munoz2009radial}. As dust is predominantly located in the plane of the disk, the bulk of the SNe Ia with blue colours likely sit in front of the central plane (from our point-of-view), experiencing minimal dust extinction, while SNe Ia within/behind the central plane will suffer more dust extinction, leading to redder SNe Ia. This picture is consistent with the observed relative number of red/blue SNe Ia, as the SNe Ia residing behind the plane will be harder to detect and hence have lower numbers. Curiously, SNe Ia located very close to the centre ($\mu < 19.5$ mag/arcsec$^2$), appear to break away from the reddening trend and cluster closely to SALT2 $c \sim 0$. Central bulges are old 3D stellar objects, with relatively little dust. However, dust is still present in the disk that cuts through the bulge. Consequently, any SNe Ia reddened by dust must reside relatively close to the centre of the plane, where the SNe Ia will almost certainly be obscured by the very bright surrounding bulge. Hence, only SNe Ia that lie sufficiently in front of the plane (away from the majority of the dust) will be visible at the bright central regions of galaxy bulges. 

If this colour trend were due to age or metallicity, we would expect the innermost SNe Ia to follow the overall trend and be the reddest, which is not seen. The lack of red SNe Ia in the bright centres of disk galaxies disfavours age or metallicity as an explanation for the apparent reddening of SN Ia colours. This is further supported by the right panels in Fig.~\ref{fig:sn_col_c}, which shows almost no correlation between SALT2 $c$ and galaxy colour (which was demonstrated to be a good tracer of age and metallicity). In fact, the only apparent observation in Fig.~\ref{fig:sn_col_c}; that red SNe Ia are missing from the bluest disk galaxies, can be explained by dust extinction, as the bluest disk galaxies are in general very young and hence have less dust content. Additionally, from the bottom-right panel of Fig.~\ref{fig:sn_sb}, at the same surface brightness values in elliptical galaxies ($\mu < 19.5$ mag/arcsec$^2$), many red normal SNe Ia are observed. For SNe with $c>0.035$ in disk containing galaxies, there are zero counts with $\mu < 19.5$ mag/arcsec$^2$ and 54 counts with $\mu > 19.5$ mag/arcsec$^2$. Whereas for SNe with $c>0.035$ in elliptical galaxies, there are 21 counts with $\mu < 19.5$ mag/arcsec$^2$ and 35 counts with $\mu > 19.5$ mag/arcsec$^2$. This strongly disfavours any observational argument based on redder normal SNe Ia being intrinsically less luminous and hence harder to observe in bright galaxy regions. 

These arguments result in the conclusion that the majority of normal SNe Ia in disk galaxies, have intrinsic SALT2 $c \lesssim 0.035$, with anything redder being due to dust. However, this can not be stated with significant certainty, as it is not known how the different progenitor channels affect this picture.  Elliptical galaxies, similarly to bulges, generally have relatively little dust and hence we would not expect any significant dust extinction. The significant presence of red  (SALT2 $c > 0.035$) normal SNe Ia in elliptical galaxies (at all surface brightness values), could point to differing progenitor channels between Elliptical and Disk galaxies, which is also supported by the significantly different $x_1$ distributions. 

\subsection{Host galaxies of `91bg-like' and `91T-like' SNe Ia}
The preferred galaxy locations of `91T-like' and `91bg-like' SNe Ia in terms of their $g-r$ SN-component-colour and host stellar mass (shown in Fig.~\ref{fig:sn_mass}) is similar to the $u-r$ galaxy colour (estimated from 25 mag arcsec$^2$ elliptical apertures) against stellar mass of \cite{SNeVII}. They found the same tendency for `91bg-like' SNe Ia to reside in the Red Sequence, while `91T-like' SNe Ia are much less likely to occur in the reddest and highest mass hosts.  \cite{SNeVI} showed that within the Red Sequence, the `91bg-like' SNe Ia have consistent host masses with the normal SNe Ia but, on average, have redder host galaxies (p-value of 0.047). This result is statistically significant  (3.5$\sigma$) in our analysis, as seen in the bottom panel of Fig.~\ref{fig:sn_mass}. The `91bg-like' events have  $g-r$ SN-component-colour that is  $0.035\pm0.01$ mag redder than the normal SNe Ia in elliptical galaxies. This colour offset, the consistency (0.5$\sigma$) of the slopes of the linear regressions for the normal and ‘91bg-like’  SNe Ia in the red sequence as well as the ‘91bg-like’ distributions in the left-hand plots of Fig.~\ref{fig:sn_col}, show that the difference in host-environments between normal and ‘91bg-like’ SNe Ia is much better captured by the galaxy colour rather than the host mass. However, it should be noted that these results are dependent on the agglomerative clustering that was used to find the red sequence.

\subsection{Benefits of disk/bulge/bar decomposition method}
In contemporary galaxy studies, the $u-r$ galaxy colour is usually preferred to $g-r$ colour  due to its ability to distinguish between young and old stellar populations \citep[e.g.~][]{galzoo}. However, as seen in Fig.~\ref{fig:sn_mass} using the $g-r$ SN-component-colour (defined in Section~\ref{sec:scc}), recovers similar galaxy correlations without relying on UV photometry. This may be of importance, as all-sky optical photometry, such as $g-$ and $r$-bands, are much more readily available (and to greater observing depths) than the corresponding UV bands.  The benefit of the SN component colour over a standard global colour, is that it removes the red bias of central bulges from the estimated global colours of disks, allowing for greater contrast between young and old stellar population (or star-forming/non star-forming regions). 

A shortcoming of some SN Ia galaxy studies is that they separate young from old stellar populations by global properties (e.g.~global colour, host stellar mass, global star formation rate, Hubble morphology). This separation is problematic as SNe Ia originating from the bulges and bars of disk galaxies are grouped with SNe Ia from young star-forming populations, even though they contain an older population of stars, much closer to those found in elliptical galaxies \citep{kauffmann1996age}. We overcome this issue by using a decomposition of the disk, bulge and bar (where required), allowing for easy separation between disk and bulge SNe Ia.

The results from this paper that were previously not seen in the literature likely highlight the importance of galaxy modelling as well as having a clean sample of host galaxies, when studying host properties. This study removes merging galaxies, edge-on disk systems, galaxies with intervening Milky Way stars, etc. These types of systems are difficult to accurately measure host stellar mass for and likely only introduce scatter in the relations between host stellar mass and $x_1$ for example. Therefore, it can be argued that SNe Ia with these types of host systems, should be removed when host environmental properties (e.g. mass-step) are being used to standardise SNe Ia for use in cosmology. 

While it is clear that there is great benefit from galaxy decomposition, the technique used in this paper does not fully reproduce a representative sample of host galaxies. In particular, low mass galaxies, asymmetric irregular galaxies, very prominent spiral arms and edge-on disk systems are underrepresented in our final sample due to being very difficult to model. Future studies could make use of more advanced decomposition codes that provide more freedom when fitting galaxies, however the more advanced the decomposition code is, the more difficult it will become to automate the procedure for a large sample of host galaxies. This is of importance as ZTF and other future sky surveys are expected to detected tens of thousands of new SNe Ia in the coming decade.

\section{Conclusions}
In this work, we performed 2D host image decomposition for SN Ia host galaxies in the ZTF DR2. Using four surface brightness models: `Elliptical', `Bulge+Disk', `Bulge+Bar+Disk' and `Disk-only', we successfully modelled 719 galaxies which were used to investigate the correlation between SN Ia photometric properties (using the SALT2 light-curve fitter) and their host environment. Our main results are as follows:
\begin{enumerate}
    \item We demonstrate that the previously observed bimodal SALT2 $x_1$ distribution seen in the literature, can be seen in the morphological galaxy type. Normal SNe Ia residing in galaxy disks are predominantly contained within the high-stretch mode, while SNe Ia residing in elliptical galaxies exist across all stretch values but constitute the majority of the low-stretch mode. We also show that SNe Ia originating from the central bulges of disk galaxies, have lower  $x_1$ values than SNe Ia in disks (4.8$\sigma$), and are consistent with the elliptical population of SNe Ia. This provides further evidence that the low stretch SNe Ia originate from older environments.
    
    \item We identify for the first time, that SNe Ia in elliptical galaxies exhibit a strong linear correlation between $x_1$ and the model derived ($g-r$) galaxy colour (`SN-component-colour') used in this study, whereby SNe Ia with lower $x_1$ values are found in redder elliptical galaxies (16.8$\sigma$, R$^2=0.37$). If a double degenerate white-dwarf progenitor model is assumed, then this relation may be driven by age rather than metallicity and that normal SNe Ia in elliptical galaxies may originate from a single progenitor channel.

    \item We identify a weaker, but still significant linear correlation between $x_1$ and the model derived ($g-r$) galaxy colour for SNe Ia in disks (5.1$\sigma$, R$^2=0.15$). While this trend also has lower $x_1$ values in redder disk, the overall slope of the relation is fundamentally different from that seen in the elliptical population of SNe Ia. This trend is likely due to a combination of age and metallicity, but is complicated by the potential diverse progenitor channels that can exist for high-stretch SN Ia.
    
    \item By taking the intersection point between the elliptical and disk linear trends, we estimate an approximate cross-over point between low- and high-stretch SNe Ia at  $x_1=-0.39\pm0.06$. This value is less negative but still consistent with the split point in the ZTF companion papers: \customcitecolor{blue}{Ginolin2024a} and \customcitecolor{blue}{Deckers2024}. However, the exact position to choose is uncertain. 

    \item We identify for the first time, that normal SNe Ia in disk containing galaxies, have a significant correlation with the local $r$-band surface brightness, whereby SNe Ia at brighter surface brightness (and hence closer to the centre) have lower $x_1$ values (6.1$\sigma$, R$^2=0.09$). This relation is consistent with the known metallicity and age gradients that exist in disk galaxies, and provides further evidence that SNe Ia in older and more metal rich environments have lower $x_1$ values. No surface brightness trend is seen in the elliptical galaxy population, further highlighting the importance of separating disk SNe Ia from those in elliptical galaxies. 

    \item As seen previously in the literature, SN Ia colour (SALT2 $c$) shows much less correlation with host environment. No significant relation was found between galaxy colour and SALT2 $c$. When analysing  $c$ versus local surface brightness, a potential dust effect may have been identified. For SNe Ia in disk containing galaxies, there is a linear correlation (3.3$\sigma$) for SNe Ia with $c > 0.035$, where redder SN Ia are found in brighter regions (more dust closer to the centre). This suggests that normal SNe Ia in disk containing galaxies have intrinsic colours of $c < 0.035$. This result can not be concluded confidently, however, as it is still unclear how different progenitor channels play a role in this picture.
    
    \item We show that the 91bg-like SNe Ia reside entirely in older populations of stars, i.e. in elliptical galaxies and in the central bulge of disk galaxies. Within the classical Red Sequence, for fixed host stellar mass, the host galaxies of 91bg-like SNe Ia were found to be $0.035\pm0.01$ mag redder than the normal SNe Ia, suggesting that the underlying progenitor system of 91bg-like SNe Ia are on average, older than those of normal SNe Ia in the Red Sequence. 

\end{enumerate}
This work has shown that by 2D modelling the host galaxies of SNe Ia and separating out different environmental types, many strong environmental correlations are recovered. In the current era of precision cosmology, we conclude that image decomposition techniques, such as the one used in this study, could better uncover the dependence of host environment on SN Ia properties, allowing for better standardisation of SN Ia luminosities. While this study has developed an image decomposition method which is fully automated and hence can be scaled to larger sample sizes, the primary shortcoming is the loss of a large fraction of the starting sample as well as a bias against low mass, irregular and merging galaxies and the exclusion of edge-on systems. This should be addressed by future works using more sophisticated decomposition programs.

\begin{acknowledgements}
RS, KM, UB, GD, MD, and JHT are supported by the H2020 European Research Council (ERC) grant no.~758638. LG acknowledges financial support from AGAUR, CSIC, MCIN and AEI 10.13039/501100011033 under projects PID2020-115253GA-I00, PIE 20215AT016, CEX2020-001058-M, and 2021-SGR-01270. This project has received funding from ERC grant no.~759194 - USNAC. LH is funded by the Irish Research Council under grant number GOIPG/2020/1387. This work has been supported by the research project grant “Understanding the Dynamic Universe” funded by the Knut and Alice Wallenberg Foundation under Dnr KAW 2018.0067 and the {\em Vetenskapsr\aa det}, the Swedish Research Council, project 2020-03444. Y-LK has received funding from the Science and Technology Facilities Council [grant number ST/V000713/1].

Based on observations obtained with the Samuel Oschin Telescope 48-inch and the 60-inch Telescope at the Palomar Observatory as part of the Zwicky Transient Facility project. ZTF is supported by the National Science Foundation under Grant No. AST-1440341 and a collaboration including Caltech, IPAC, the Weizmann Institute of Science, the Oskar Klein Center at Stockholm University, the University of Maryland, the University of Washington, Deutsches Elektronen-Synchrotron and Humboldt University, Los Alamos National Laboratories, the TANGO Consortium of Taiwan, the University of Wisconsin at Milwaukee, and Lawrence Berkeley National Laboratories. Operations are conducted by COO, IPAC, and UW. The ZTF forced-photometry service was funded under the Heising-Simons Foundation grant \#12540303 (PI: Graham). The Gordon and Betty Moore Foundation, through both the Data-Driven Investigator Program and a dedicated grant, provided critical funding for SkyPortal.
The Legacy Surveys consist of three individual and complementary projects: the Dark Energy Camera Legacy Survey (DECaLS; Proposal ID \#2014B-0404; PIs: David Schlegel and Arjun Dey), the Beijing-Arizona Sky Survey (BASS; NOAO Prop. ID \#2015A-0801; PIs: Zhou Xu and Xiaohui Fan), and the Mayall z-band Legacy Survey (MzLS; Prop. ID \#2016A-0453; PI: Arjun Dey). DECaLS, BASS and MzLS together include data obtained, respectively, at the Blanco telescope, Cerro Tololo Inter-American Observatory, NSF’s NOIRLab; the Bok telescope, Steward Observatory, University of Arizona; and the Mayall telescope, Kitt Peak National Observatory, NOIRLab. Pipeline processing and analyses of the data were supported by NOIRLab and the Lawrence Berkeley National Laboratory (LBNL). The Legacy Surveys project is honored to be permitted to conduct astronomical research on Iolkam Du’ag (Kitt Peak), a mountain with particular significance to the Tohono O’odham Nation.
NOIRLab is operated by the Association of Universities for Research in Astronomy (AURA) under a cooperative agreement with the National Science Foundation. LBNL is managed by the Regents of the University of California under contract to the U.S. Department of Energy.

This project used data obtained with the Dark Energy Camera (DECam), which was constructed by the Dark Energy Survey (DES) collaboration. Funding for the DES Projects has been provided by the U.S. Department of Energy, the U.S. National Science Foundation, the Ministry of Science and Education of Spain, the Science and Technology Facilities Council of the United Kingdom, the Higher Education Funding Council for England, the National Center for Supercomputing Applications at the University of Illinois at Urbana-Champaign, the Kavli Institute of Cosmological Physics at the University of Chicago, Center for Cosmology and Astro-Particle Physics at the Ohio State University, the Mitchell Institute for Fundamental Physics and Astronomy at Texas A\&M University, Financiadora de Estudos e Projetos, Fundacao Carlos Chagas Filho de Amparo, Financiadora de Estudos e Projetos, Fundacao Carlos Chagas Filho de Amparo a Pesquisa do Estado do Rio de Janeiro, Conselho Nacional de Desenvolvimento Cientifico e Tecnologico and the Ministerio da Ciencia, Tecnologia e Inovacao, the Deutsche Forschungsgemeinschaft and the Collaborating Institutions in the Dark Energy Survey. The Collaborating Institutions are Argonne National Laboratory, the University of California at Santa Cruz, the University of Cambridge, Centro de Investigaciones Energeticas, Medioambientales y Tecnologicas-Madrid, the University of Chicago, University College London, the DES-Brazil Consortium, the University of Edinburgh, the Eidgenossische Technische Hochschule (ETH) Zurich, Fermi National Accelerator Laboratory, the University of Illinois at Urbana-Champaign, the Institut de Ciencies de l’Espai (IEEC/CSIC), the Institut de Fisica d’Altes Energies, Lawrence Berkeley National Laboratory, the Ludwig Maximilians Universitat Munchen and the associated Excellence Cluster Universe, the University of Michigan, NSF’s NOIRLab, the University of Nottingham, the Ohio State University, the University of Pennsylvania, the University of Portsmouth, SLAC National Accelerator Laboratory, Stanford University, the University of Sussex, and Texas A\&M University.
BASS is a key project of the Telescope Access Program (TAP), which has been funded by the National Astronomical Observatories of China, the Chinese Academy of Sciences (the Strategic Priority Research Program “The Emergence of Cosmological Structures” Grant \# XDB09000000), and the Special Fund for Astronomy from the Ministry of Finance. The BASS is also supported by the External Cooperation Program of Chinese Academy of Sciences (Grant \# 114A11KYSB20160057), and Chinese National Natural Science Foundation (Grant \# 12120101003, \# 11433005).
The Legacy Survey team makes use of data products from the Near-Earth Object Wide-field Infrared Survey Explorer (NEOWISE), which is a project of the Jet Propulsion Laboratory/California Institute of Technology. NEOWISE is funded by the National Aeronautics and Space Administration.
The Legacy Surveys imaging of the DESI footprint is supported by the Director, Office of Science, Office of High Energy Physics of the U.S. Department of Energy under Contract No. DE-AC02-05CH1123, by the National Energy Research Scientific Computing Center, a DOE Office of Science User Facility under the same contract; and by the U.S. National Science Foundation, Division of Astronomical Sciences under Contract No. AST-0950945 to NOAO.

\end{acknowledgements}

\bibliographystyle{aa}
\bibliography{references} 

\end{document}